\documentclass[onecolumn%
 reprint,
 amsmath,amssymb,
]{revtex4-2}

\usepackage{graphicx}
\usepackage{dcolumn}
\usepackage{bm}
\usepackage{multirow}
\usepackage{array}
\usepackage{appendix}
\usepackage{hyperref}

\begin{document}

\preprint{APS/123-QED}

\title{A Magnetothermodynamic Theory of Energy Transport In Collisionless Plasmas}

\author{Dominic Payne}
 \affiliation{Climate and Space Sciences and Engineering, University of Michigan, Ann Arbor, MI.}


\begin{abstract}
Presented here is an exploratory conceptual framework with the aim of describing electromagnetic and thermal energy transport and transfer in collisionless plasmas in a coherent way, starting from an 'arrow-of-time' perspective.  That is, arguments about the transport of energy are fundamentally based on the idea that systems tend to fill the phase space available to them.  First, we describe electric fields, magnetic fields, and plasma particles as interacting 'subsystems' that are thermodynamically linked in specific ways.  Second, we discuss the importance of the stochastic electric field in the transport of plasma particles down energy density gradients until the energy flux contributions from so-called electrothermodynamic (ETD) interactions become sufficiently uniform.  Third, the logic of ETD transport is extended to a magnetized system invoking 2 types of magnetothermodynamic (MTD) transport: 1) transport of electromagnetic energy through stochastic Poynting flux and 2) plasma energy transport through stochastic 'drift' energy fluxes associated with local $E \times B$ drift motion. The magnitudes of the ETD and MTD mechanisms are tied to local characteristic scales of the electric field fluctuations and  critical scales are derived beyond which ETD mechanisms dominate MTD mechanisms.  Finally, we discuss the interplay between electrothermodynamic and magnetothermodynamic equilibria (ETE and MTE), dynamics in the small critical scale (MTD) limit, and the potential relevance of this type of description to the problem of reconnection onset and scale coupling.
\end{abstract}

\maketitle

\section{\label{sec:level1} Introduction}
The mid-19th century saw the rise of modern thermodynamic theory with statistical-mechanical explanations underpinning the concepts of irreversibility and entropy \cite{clausius1850ueber,boltzmann2003relation,carnot1824reflections}.  These concepts are unique in classical physics because they provide a probabilistic explanation for emergent macroscopic behavior in systems with many interacting particles. This emergent behavior suggests the existence of a preferential 'arrow of time' for closed macroscopic systems, which appears to contradict the time-symmetric physics of small-scale kinetic interactions.  Particle collisions are typically the mechanism by which classical systems irreversibly distribute their thermal energy.  Collisionless plasmas, however, have no equivalent dissipative mechanism and the evolution of phase space is dominated by electromagnetic forces rather than internal dissipation.  Due to the lack of a strong thermalization mechanism, such systems can sustain highly non-Maxwellian velocity distribution functions (VDFs) \cite{hesse2014electron,shuster2014highly,bessho2014electron} and are typically well-described by the Vlasov-Maxwell equations, which are fully reversible in principle.  In practice,  collisionless plasma processes such as magnetic reconnection and turbulence do not appear to be reversible\cite{xuan2021reversibility} and collisionless dissipation still statistically resembles collisional dissipation \cite{bandyopadhyay2023collisional}.   

Expansions on kinetic and fluid theories of collisionless plasma thermodynamics include different measures of kinetic entropy \cite{liang2020kinetic, argall2022theory},  generalized higher-order moment thermodynamic approaches \cite{cassak2023quantifying,barbhuiya2024higher}, pressure-strain interaction term \cite{yang2022pressure} and its contributions \cite{cassak2022pressure1,cassak2022pressure2} representing the conversion of bulk to thermal energy and magnetic analogues to pressure-strain \cite{barbhuiya2024magnetic}.  There has also been work on the interplay between kinetic-scale dynamics and the establishment of novel types of equilibria unique to collisionless plasmas. Examples include an explanation for the kinetic stabilization of temperature gradients in electron-scale current sheets \cite{shuster2021electron}, equilibrium selection mechanisms in mixed-equilibrium current sheets \cite{yoon2023equilibrium} and kinetic drivers of MHD equilibration processes \cite{park2025kinetic}.  

The multiscale nature of magnetic reconnection invites a more unified description of collisionless energy transport.  Significant advances in our understanding of magnetic reconnection have been inferred from analyses of the energy flux budgets\cite{birn2005energy,birn2010energy,fargette2024statistical,eastwood2026magnetic}.  Competing energy flux structures have been invoked to explain various aspects of reconnection including theories of the reconnection rate \cite{liu2022first,liu2025analytical}, 'magnetic slippage' mechanisms based on magnetic flux transport (MFT) \cite{li2021identification,qi2022magnetic,li2025magnetic}, and energy transport mechanisms governing the temporal evolution of x-line structures \cite{genestreti2018assessing,payne2020energy,payne2021origin,payne2025situ}.  

Collisionless plasma turbulence is another multiscale process responsible for energy dissipation.  In fact, the separate categorization of reconnection-driven vs turbulence-driven energy dissipation is not straightforward because the characteristics of energy transfer in reconnection and turbulence exhibit structural similarities \cite{adhikari2021energy} and both processes are mutually influential \cite{stawarz2024interplay}.  Recent results \cite{khan2026factors} suggest that reconnection rates within turbulent environments may be controlled by the global turbulent magnetic field at the correlation scale, further complicating the relationship between local and global processes while demonstrating a fundamental connection between energy conversion via reconnection and turbulent energy dissipation.  

The following sections propose a theoretical description which may be useful for describing and understanding collisionless energy transport processes.  Section \ref{sec:systems} establishes the constraints on any collisionless system imposed by energy conservation and continuity.  Section \ref{sec:scales} goes deeper into the role of stochastic electric field fluctuations and the scale-dependent energy transport mechanisms they create.  Section \ref{sec:emergent} expands on the arguments of those previous to describe emergent, macroscopic energy transport characteristics and defines different types of energy equilibria.  Section \ref{sec:reconnection} brings everything together to describe reconnection onset using the principles established in sections \ref{sec:systems}-\ref{sec:emergent}.  In section \ref{sec:discussion}, the main pillars of this theory are summarized and the implications of and complications to it are considered.  Section \ref{sec:conclusion} briefly summarizes the main conclusions.

\section{\label{sec:systems} Systems and Subsystems}
\subsection{Global Energy Conservation}
\begin{figure}[h]
    \includegraphics[scale=0.5]{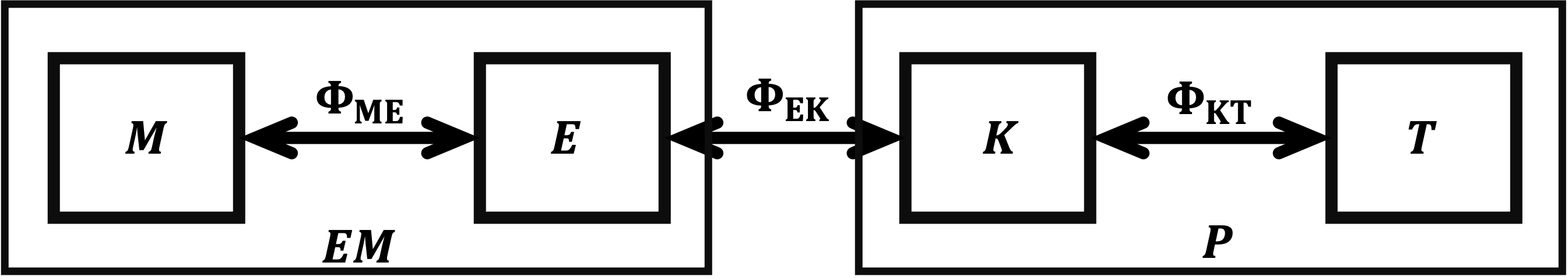}
    \caption{The magnetic (M), electric (E), and plasma (P) subsystems are the three regimes in which energy can exist.  Because magnetic fields cannot directly do work on particles at a kinetic level, the M and P subsystems cannot directly exchange energy, so any transfer of magnetic energy to particle energy must be mediated by the E subsystem.  }
    \label{fig:globalsystems}
\end{figure}

The energy content of an isolated system is conserved, but it may exist in different forms.  In a collisionless plasma, energy must be contained in either the electromagnetic fields ($EM$), or the plasma particles ($P$), which will be referred to as `subsystems' that can exchange energy.  These subsystems can be divided into their own set of subsystems, where $EM$ can contain energy in the magnetic ($M$) or electric ($E$) fields and $P$ can contain bulk kinetic ($K$) and thermal ($T$) contributions.  Global energy conservation dictates that the total energy of all subsystems must remain constant: 
\begin{equation}
\partial_t (U_{_{EM}}+ U_{_P}) = \partial_t (U_{_M}+ U_{_E}+ U_{_K} + U_{_T}) = 0 \label{eqn:encon1}
\end{equation}
Equation \ref{eqn:encon1} allows for energy transfer between subsystems so long as the total energy remains constant.  While magnetic energy can ultimately be converted to plasma energy during magnetic reconnection or plasma turbulence, the only kinetic mechanism that can do work to a charged particle is direct acceleration by an electric field.  Therefore any transfer of energy from $M$ to $P$ is in reality a transfer of energy from $M$ to $E$ to $P$ ($M\rightarrow E\rightarrow P$), where E is the intermediary between $M$ and $P$.  The only part of P directly connected to $E$ is $K$ ($E\leftrightarrow K$), since direct acceleration immediately provides bulk energy which can dissipate into thermal energy ($K\rightarrow T$) over time.  These mechanistic constraints are visualized in \ref{fig:globalsystems}, where the connections represent possible pathways of energy transfer.  The time evolution of the total energy in each subsystem can be represented by one or a combination of the $\Phi$ terms which represent energy transport between subsystems.
\begin{align}
  \partial_tU_{_{M}} &= - \Phi_{_{ME}} \label{eqn:phi1}\\
   \partial_tU_{_{E}} &= \Phi_{_{ME}} - \Phi_{_{EK}}\label{eqn:phi2} \\
   \partial_tU_{_{K}} &= \Phi_{_{EK}} - \Phi_{_{KT}} \\
   \partial_tU_{_{T}} &= \Phi_{_{KT}} \label{eqn:phi4}
\end{align}  
where $\Phi_{_{A,B}}$ indicates a transport of energy from subsystem $A$ to $B$ ($A\rightarrow B$).  The mechanistic constraints on energy transport between subsystems do not rule out correlated changes in disconnected subsystems.  It is possible for $\partial_t U_{_M} \simeq -\partial_t U_{
_T}$,  so long as $\Phi_{_{ME}} \simeq \Phi_{_{KT}}$, which from equation \ref{eqn:phi2} gives  $\partial_t U_{_E} \simeq \Phi_{_{KT}} - \Phi_{_{EK}} = -\partial_t U_{_K}$.  The condition of $\partial_tU_{_M} \simeq -\partial_tU_{_T}$ therefore automatically implies an inverse relationship between the $E$ and $K$ subsystems $\partial_tU_{_E} \simeq -\partial_tU_{_K}$.  In the simplest version of this condition, all $\Phi$ terms are equal and $\partial_tU_{_E} = \partial_tU_{_K} = 0$, which implies energy transfer from $M\rightarrow T$ where the energy moves through subsystems $E$ and $K$ without any net accumulation or depletion of $U_{_E}$ and $U_{_K}$.  The $E$ and $K$ subsystems in such a scenario merely act as conduits through which energy is transported between $M$ and $T$.       

The $\Phi$ terms represent the power associated with the energy transfer between subsystems globally.  Equations \ref{eqn:phi1}-\ref{eqn:phi4} are straightforward because there are no external sources of energy that contribute to changes in the total energy of the system, so any loss of energy in one subsystem must result in energy gain in another.  Where this logical framework of energy conduits between subsystems becomes important and potentially more useful is in the treatment of \textit{local} energy conservation for a local system that is free to interact with its surroundings.

\subsection{Local Energy Continuity}
\begin{figure}[h]
\includegraphics[scale=0.5]{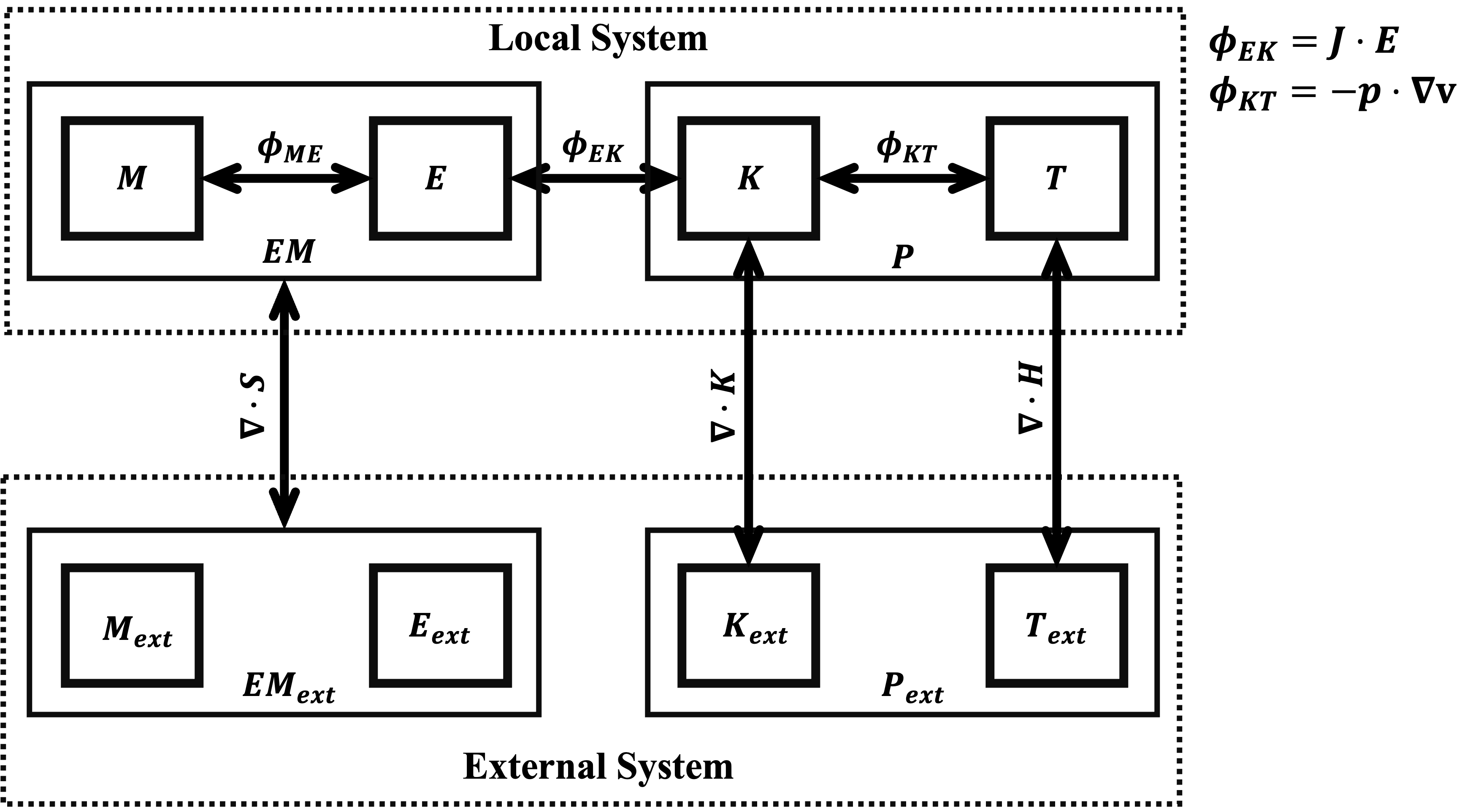}
\caption{System diagram depicting the network of connections between subsystems}
\label{fig:localsystems}
\end{figure}

An expression for the \textit{local} conservation of energy in an arbitrary region in space takes the form of a basic energy continuity equation in which the evolution of the total \textit{energy density} must be balanced by the total amount of energy flux flowing across the boundaries of the region.  For an electric field $\vec{E}$, magnetic field $\vec{B}$, plasma density $n$, temperature $T$ and velocity $\vec{v}$, the associated energy densities and fluxes are defined \cite{birn2005energy} as: 
\begin{equation}\label{eqn:endensities}
u_{_{EM}} = u_{_M} + u_{_E}  = \frac{|\vec{B}|^2}{2\mu_0} +\frac{\epsilon_0E^2}{2} \ \ \bigg| \ \ u_{_P} = u_{_K} + u_{_T} = \frac{nm|\vec{v}|^2}{2} + \frac{3}{2}nk_bT
\end{equation}
\begin{equation}
\vec{S} = \frac{\vec{E}\times \vec{B}}{\mu_0}  \ \ \bigg| \ \ \vec{K} = \frac{nm\vec{v}^3}{2} \ \ \bigg| \ \ \vec{H} = \frac{5}{2}nk_bT \ \vec{v}
\end{equation}
\begin{equation}
\partial_t  (u_{_M}+u_{_E}+u_{_K} + u_{_T}) = - \nabla\cdot(\vec{S} + \vec{K} + \vec{H})\label{eqn:endencon0}
\end{equation}
where $u_{_M}$, $u_{_E}$, $u_{_K}$, and $u_{_T}$  represent the magnetic, electric, bulk kinetic, and thermal energy density, respectively.  $u_{_P}$ includes a contribution from bulk flow energy $u_{_K}$ and from thermal energy $u_{_T}$.  $\vec{S}$, $\vec{K}$, and $\vec{H}$ represent the electromagnetic energy flux (Poynting flux), bulk kinetic and enthalpy fluxes, respectively.  $\epsilon_0$, $\mu_0$, and $k_b$ are the permittivity and permeability of free space and the Boltzmann constant, respectively.  The constraints on the local evolution of the energy densities can take a similar form as the global energy constraints in equations \ref{eqn:phi1}-\ref{eqn:phi4}, using lowercase $\phi_{_{AB}}$ to represent the power densities of energy transfer from subsystem $A$ to $B$ and introducing energy flux divergence terms which connect the local subsystems with the surroundings.  For simplicity, the time evolution of $u_{_E}$ and $u_{_M}$ are not expressed separately and there is no $\phi_{_{ME}}$ in equation \ref{eqn:phi5}.  However, the separate $M$ and $E$ subsystems and $\phi_{_{ME}}$ are still shown in the full system diagram (figure \ref{fig:localsystems}).
\begin{align}
  \partial_tu_{_{EM}} &= - \phi_{_{EK}} - \nabla\cdot\vec{S}\label{eqn:phi5}\\
   \partial_tu_{_{K}} &= \phi_{_{EK}} - \phi_{_{KT}} - \nabla \cdot \vec{K}  \\
   \partial_tu_{_{T}} &= \phi_{_{KT}}- \nabla \cdot \vec{H}\label{eqn:phi8}
\end{align}
Unlike the global expressions, equations \ref{eqn:phi5}-\ref{eqn:phi8} describe a system open to its surroundings.  The net transport of electromagnetic, kinetic, and thermal energy between the external and local systems is represented by the $\nabla\cdot\vec{S}$, $\nabla\cdot\vec{K}$, and $\nabla\cdot\vec{H}$ terms, respectively.  

Some of the $\phi$ terms can be specified.  The conversion of electric energy to bulk flow energy ($E\rightarrow K$) is $\phi_{_{EK}} = \vec{J}\cdot\vec{E}$ where $\vec{J}$ is the current density and the conversion of bulk flow energy to thermal energy ($K\rightarrow T$) is given by the pressure-strain interaction term $\phi_{_{KT}} = -\overleftrightarrow{p}\cdot\nabla\vec{v}$, where $\overleftrightarrow{p}$ is the pressure tensor \cite{matthaeus2020pathways, yang2022pressure}.  Simplified expressions for $\partial_t u_{_{EM}}$ and $\partial_t u_{_{P}}$ are presented in equations \ref{eqn:endencon1} (Poynting's Theorem) and \ref{eqn:endencon2}, respectively, including $\vec{J}\cdot\vec{E}$ representing the energy transport between the EM and P subsystems ($EM \leftrightarrow P$).
\begin{align}
\partial_t u_{_{EM}} &= -\vec{J}\cdot\vec{E} - \nabla \cdot \vec{S} \label{eqn:endencon1} \\
\partial_t u_{_P} &=  \vec{J}\cdot\vec{E} - \nabla \cdot (\vec{K} + \vec{H})\label{eqn:endencon2}
\end{align}
where $\vec{J}$ is the current density and $\vec{J}\cdot\vec{E}$ represents power density associated with the transfer of energy from electromagnetic fields to particles.  Equations \ref{eqn:phi5}-\ref{eqn:endencon2} account for energy evolution in time, transport through space, and transfer between subsystems.

\section{\label{sec:scales} Scale-Dependent Energy Transport}
The terms that couple subsystems in section \ref{sec:systems} constrain how energy can move through the system, but energy continuity alone cannot determine which terms dominate in a given scenario, nor can it establish any preferred evolution of the system in time.  This section explores how the various energy transport and transfer mechanisms of section \ref{sec:systems} depend on the characteristic spatiotemporal scales of the system's electric fields.  

\subsection{The Role of $\delta \vec{E}$ Fluctuations}\label{sec:etd}
Consider a particle initially at rest, then accelerated by a local electric field of magnitude $\delta E$ over a sufficiently small temporal ($\tau$) and spatial ($\ell$) scale such that $\delta E(r_0,t_0) \sim \delta E(r_0+\ell,t_0 + \tau)$.  The interaction moves the particle in phase space by $\ell$ and $\delta v$.   The work done by the electric field during this interaction is given by
\begin{equation}
\delta W = q\delta E \ell 
\end{equation}

By equating $\delta W$ with the change in kinetic energy, one can solve for the change in velocity-space coordinate 
\begin{equation}\label{eqn:dvel}
\delta v_{_E}= \sqrt{\frac{2q\delta E\ell}{m}}
\end{equation}

This represents the maximum possible velocity-space displacement for a charged particle of mass $m$ due to the influence of $\delta E$ of length scale $\ell$.  Alternatively $\delta v_{_E}$ can be expressed in terms of the timescale $\tau$ of the interaction.
\begin{equation}\label{eqn:dvet}
\delta v_{_E} = \frac{q \delta E}{m} \tau
\end{equation}

The associated maximum change in kinetic energy flux density magnitude $\delta K_{_E}$ for an ensemble of particles of density $n$ undergoing the same process over the length scale $\ell$  or timescale $\tau$ is then:

\begin{equation}
\delta K_{_E} = \frac{nm\delta v_{_E}^3}{2}\label{eqn:firstdeltake}
\end{equation}

\begin{equation}\label{eqn:keterms}
\delta K_{_E} = \sqrt{\frac{2 n^2 q^3 \delta E^3}{m}}\ell^{3/2} \ \  \ \ OR  \ \ \ \ \delta K_{_E}  = \frac{nq^3 \delta E^3}{2m^2}\tau^3
\end{equation}

If the ensemble also has associated with it a nonzero temperature $T$, the net acceleration by $\delta E$ will also introduce a nonzero change in the enthalpy flux density $\delta H_{_E}$.

\begin{equation}
\delta H_{_E} = \frac{5}{2}nk_bT \ \delta v_{_E}\label{eqn:firstdeltahe}    
\end{equation}

\begin{equation}\label{eqn:heterms}
\delta H_{_E} = 5nk_bT \ \sqrt{\frac{q\delta E\ell}{2m}}\ \ \ \ OR \ \ \ \ \delta H_{_E} = \frac{5nqk_bT \delta E}{2m}\tau    
\end{equation}

There are subtle differences between $\delta K_{_E}$ and $\delta H_{_E}$.  $\delta K_{_E}$ represents the kinetic energy flux due to the gain in bulk kinetic energy by acceleration along $\delta E$.  While $\delta v_{_E}$ contributes to a nonzero $\delta H_{_E}$, the gained enthalpy flux does not represent an additional gain in the total energy of the ensemble that is not accounted for by $\delta K_{_E}$.  It only represents the transport of thermal energy which was already present before the interaction with the electric field.  The energy density of the ensemble consists of both the thermal energy density $\frac{3}{2}nk_bT$ and the energy of the bulk flow $\frac{1}{2}nmv^2$, so while the $\delta v_{_E}$ term contributes to both kinetic and thermal energy \textit{flux}, it does not contribute to thermal energy \textit{density}. 

The maximum power density associated with energy transfer to the ensemble can also be determined using equations \ref{eqn:dvel} and \ref{eqn:dvet} to obtain $\delta J_{_E}$ and express $\delta J_{_E}\delta E$ in terms of $\ell$ or $\tau$:

\begin{equation}\label{eqn:deltaj}
\delta J_{_E} = nq\delta v_{_E}
\end{equation}

\begin{equation}\label{eqn:deltaj}
\delta J_{_E}\delta E = \sqrt{\frac{2n^2q^3\delta E^3 \ell}{m}} \  \ \ \ OR \ \ \ \  \delta J_{_E}\delta E =\frac{nq^2\delta E^2\,\tau}{m} 
\end{equation}

\subsection{\label{sec:ensembles} Influence of $\delta\vec{E}\times\vec{B}$ Effects}
Now consider a  magnetic field $\vec{B}$ imposed upon the ensemble from section \ref{sec:etd}, thus producing fluctuations of Poynting flux density $\delta S$ due to the $\delta \vec{E}\times\vec{B}$ contributions.   
\begin{equation}
\delta S = \frac{\delta E \ B}{\mu_0}\label{eqn:sterm}
\end{equation}
The $\delta \vec{E}\times\vec{B}$ contributions will also produce fluctuations in the particle drift velocity $\delta v_{_D}$ thereby introducing particle energy transport due to drift motion in addition to the electromagnetic energy transport from Poynting flux.  $\delta v_{_D}$ and its contribution to the kinetic and enthalpy flux ($\delta K_{_D}$ and $\delta H_{_D}$, respectively) are expressed below  
\begin{align}
\delta v_{_D} &= \frac{\delta E \ B}{B^2} = \frac{\delta E}{B}\label{eqn:vdterm} \\
\delta K_{_D} &= \frac{nm\delta v_{_D}^3}{2} = \frac{nm}{2}\bigg(\frac{\delta E}{B}\bigg)^3\label{eqn:kdterm} \\
\delta H_{_D} &= \frac{5}{2}nk_bT\delta v_{_D} = \frac{5}{2}nk_bT \Big(\frac{\delta E}{B}\Big)\label{eqn:hdterm} 
\end{align}

\subsection{\label{sec:critscales} Critical Scales}
The $\delta E$ fluctuations influence all the energy flux fluctuation terms, including the $\delta \vec{E}\times\vec{B}$ aligned terms in equations \ref{eqn:sterm}, \ref{eqn:kdterm} and \ref{eqn:hdterm} and the $\delta \vec{E}$ aligned terms in equations \ref{eqn:keterms} and \ref{eqn:heterms}.  The mutual dependence on $\delta E$ fluctuations implies that the energy transport mechanisms mutually influence each other.  Unlike the drift terms, the terms due to direct acceleration by $\delta\vec{E}$ are dependent on $\ell$ and $\tau$.  This built-in scale dependence determines whether energy transport mechanisms along $\delta \vec{E}$ are more significant than those along $\delta \vec{E}\times\vec{B}$.  The relevant characteristic energy flux densities in this system are listed together below, including both the $\ell$ and $\tau$-dependent versions of $\delta K_{_E}$ and $\delta H_{_E}$.

\begin{equation}
\delta K_{_E} = \sqrt{\frac{2 n^2 q^3 \delta E^3}{m}}\ell^{3/2} \ \  \ \ OR  \ \ \ \ \delta K_{_E}  = \frac{nq^3 \delta E^3}{2m^2}\tau^3\label{eqn:kterms}    
\end{equation}
\begin{equation}
\delta H_{_E} = 5nk_bT \ \sqrt{\frac{q\delta E\ell}{2m}}\ \ \ \ OR \ \ \ \ \delta H_{_E} = \frac{5nqk_bT \delta E}{2m}\tau\label{eqn:hterms}   
\end{equation}
\begin{equation}
\delta K_{_D} = \frac{nm\delta E^3}{2B^3} \ \ \bigg| \ \ \delta H_{_D} = \frac{5nk_bT \delta E}{2B} \ \ \bigg| \ \ \delta S = \frac{\delta EB}{\mu_0}\label{eqn:driftterms}
\end{equation}

\begin{figure}[h]
    \includegraphics[scale=0.5]{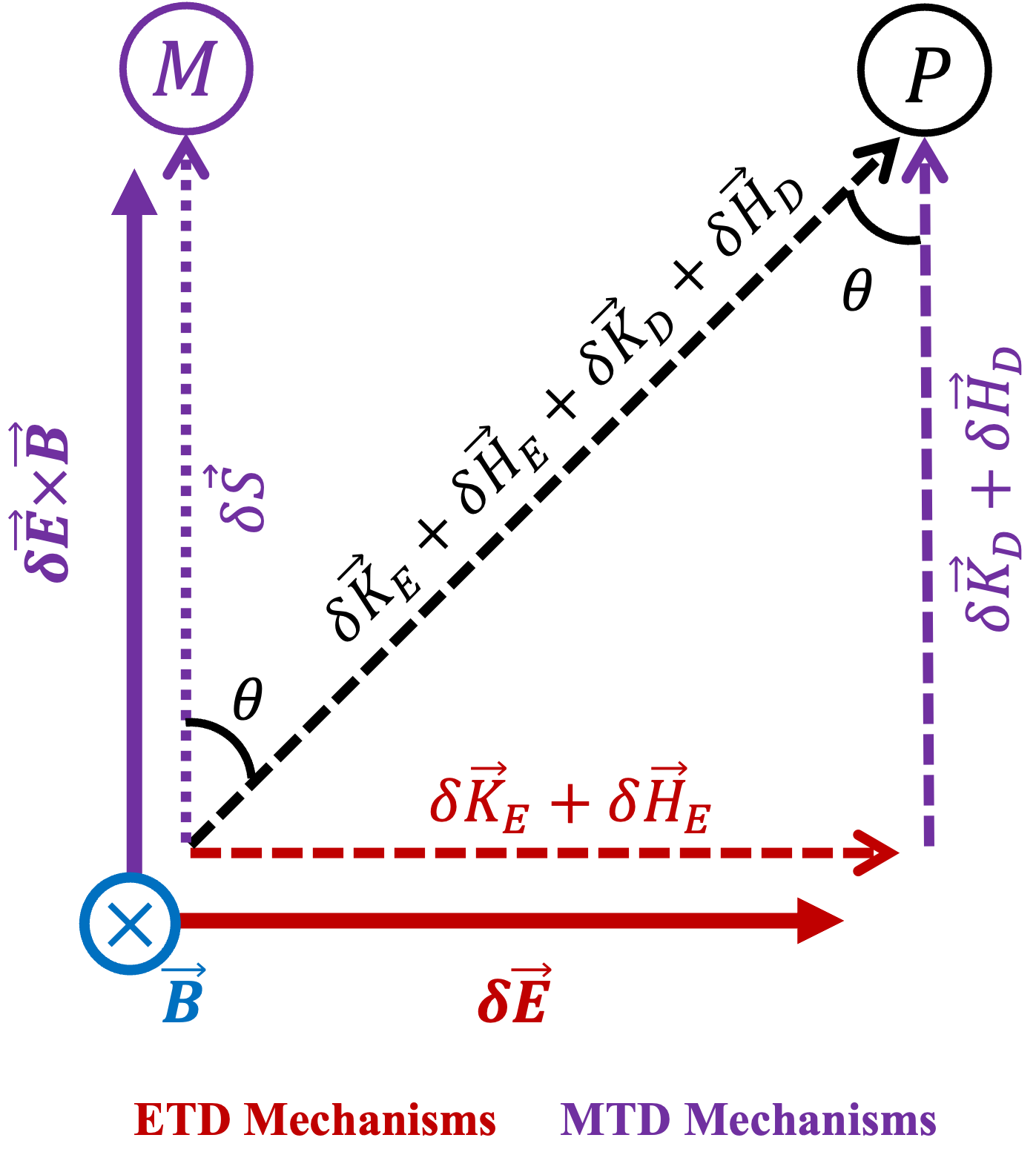}
    \caption{Contributions to plasma (P) and magnetic (M) energy transport from a $\delta E$ fluctuation}
    \label{fig:singint}
\end{figure}

$\delta K_{_E}$ represents the change in particle energy flux due to direct acceleration by $\delta E$, and therefore is the only term that represents direct energy exchange between the electric and plasma subsystems $E\leftrightarrow P$.  $\delta H_{_E}$ is also dependent on $\delta E$ and its characteristic scales, but it only represents a transport of thermal energy in space along $\delta \vec{E}$.  $\delta K_{_D}$, $\delta H_{_D}$, and $\delta S$ represent the transport of bulk, thermal, and electromagnetic energy in space due to mechanisms acting in the $\delta \vec{E} \times \vec{B}$ direction.  Throughout the remainder of this study the energy transport mechanisms in equations \ref{eqn:kterms} and \ref{eqn:hterms} will be categorized as electro-thermodynamic (ETD) mechanisms since they exist independently from any magnetic field influence, while the terms in equations \ref{eqn:driftterms} will be referred to as magneto-thermodynamic (MTD) mechanisms.  A diagram depicting the relevant energy transport contributions in the $\delta \vec{E},\vec{B}, \delta\vec{E}\times\vec{B}$ coordinate system is presented in figure \ref{fig:singint}. 

One can compare the various energy flux contributions to each other if the magnitudes of $n$,$m$,$T$,$\delta E$, and $B$ are known.  This is a straightforward task for the terms in equations \ref{eqn:driftterms}, but is not as straightforward to quantify the terms in equations \ref{eqn:kterms} and \ref{eqn:hterms} due to their dependence on arbitrary fluctuation scales $\tau$ and $\ell$.  If $\delta E$ approaches infinite length and duration,  $\ell, \tau \rightarrow\infty$ and the terms in equations \ref{eqn:kterms} and \ref{eqn:hterms} become infinitely large while those in equations \ref{eqn:driftterms} are unaffected.  Therefore, the relative energy flux contribution from $\delta E$ interactions in equations \ref{eqn:kterms} and \ref{eqn:hterms} versus the drift terms in equations \ref{eqn:driftterms} is dependent on the characteristic scales of the system's $\delta E$ fluctuations.  Set equalities between various terms in equations \ref{eqn:kterms}, \ref{eqn:hterms} and \ref{eqn:driftterms}, and solve for $\tau$ and $\ell$ to get the critical scales above which different forms of energy transfer begin to dominate.  The resulting critical scales are shown in table \ref{tab:criticalscales} and their derivations are shown in the appendix.  
\begin{table}[h]
\centering
\renewcommand{\arraystretch}{2.5}
\setlength{\tabcolsep}{18pt}
\begin{tabular}{c|c|c}
\hline\hline
\textbf{Condition} & \textbf{Critical Timescale} & \textbf{Critical Lengthscale} \\
\hline
$\delta K_{_E}=\delta H_{_E}$ & 
$\tau_{_{KH}} = \left(\dfrac{5mk_bT}{q^2\delta E^2}\right)^{1/2}$ & 
$\ell_{_{KH}} = \dfrac{5k_bT}{2q\delta E}$ \\[10pt]
\hline
$\delta K_{_E}=\delta S$ & 
$\tau_{_{KS}} = q^{-1} \left(\dfrac{2B}{\mu_0 n}\right)^{1/3}\left(\dfrac{m}{\delta E}\right)^{2/3}$ & 
$\ell_{_{KS}} = q^{-1} \left(\dfrac{B}{\mu_0 n}\right)^{2/3}\left(\dfrac{m}{2\delta E}\right)^{1/3}$ \\[10pt]
\hline
$\delta K_{_E} = \delta K_{_D}$ & 
\multirow{2}{*}{$\tau_{_D} = \dfrac{m}{qB} = 1/\omega_c$} & 
\multirow{2}{*}{$\ell_{_D} = \dfrac{m\delta E}{2qB^2}$} \\
$\delta H_{_E}=\delta H_{_D}$ & & \\[10pt]
\hline
$\delta H_{_E}=\delta S$ & 
$\tau_{_{HS}} = \left(\dfrac{2mB}{5q\mu_0 n k_b T}\right)$ & 
$\ell_{_{HS}} = \left(\frac{2m\delta E}{q}\right)\left(\dfrac{B}{5\mu_0 n k_b T}\right)^{2}$ \\[10pt]
\hline\hline
\end{tabular}
\caption{Critical timescales and lengthscales derived from ratios of energy flux density terms. Each row gives the condition under which ETD and MTD transport mechanisms are in balance, along with the corresponding characteristic scales $\tau$ and $\ell$.}
\label{tab:criticalscales}
\end{table}

The terms in table \ref{tab:criticalscales} include critical length and time scales over which the bulk energy transport exceeds the thermal energy transport (the KH scales) and scales over which ETD mechanisms exceed MTD mechanisms (the KS, D, and HS scales).  Note that the critical scale $\tau_{_D}$ is the inverse cyclotron frequency $\omega^{-1} = m/qB$, which is consistent with intuition.  If $\tau \ll \tau_{_D}$, then the electric field vector can fluctuate about zero many times before the particle completes a single orbit of gyro-motion.  The result is that  the gyro-period averaged electric field is negligible $\langle \vec{E}\rangle_{\omega_c^{-1}}\approx 0$, so there is negligible p;acceleration in the guiding-center limit and any transport of the particle's energy beyond kinetic scales is tied to the local $\vec{E}\times\vec{B}$ drift motion. If $\tau \gg \tau_{_D}$, $\langle \vec{E} \rangle_{\omega_c^{-1}} \neq 0$, which results in a net acceleration of the guiding center and makes ETD transport of particles more efficient than transport by $\vec{E}\times\vec{B}$ drift.  The intermediate condition $\tau \sim \tau_{_D}$ is essentially the condition of cyclotron resonance, where the period of $\delta E$ fluctuation matches that of the gyroperiod such that the particle gains energy without significant acceleration of its guiding center.  If $\ell \gg  \ell_{_D}$ the ETD transport dominates MTD transport of particle energy.  $\ell_{_D}$ can also be expressed in terms of the drift velocity and gyro-frequency ($l_{_D} = \delta v_{_D}/{2\omega_c}$).

The angle $\theta$ between the MTD transport and the total plasma energy transport shown in figure \ref{fig:singint} is defined by:    
\begin{equation} \label{eqn:angle}
\tan(\theta) = \frac{\delta K_{_E} + \delta H_{_E}}{\delta K_{_D} + \delta H_{_D}}
\end{equation}
In the kinetic limit, where the thermal energy is negligible compared to the bulk flow, the $\delta H$ terms can be ignored. 

If $\delta K_{_{E,D}} \gg \delta H_{_{E,D}}$ or $\tau,\ell \gg \tau_{_{KH}},\ell_{_{KH}}$:
\begin{equation} \label{eqn:anglek}
\tan(\theta) = \frac{\delta K_{_E}}{\delta K_{_D}} = \bigg(\frac{\delta v_{_E}}{\delta v_{_D}}\bigg)^3
\end{equation}

\begin{equation}
\tan(\theta) = \bigg(\frac{\tau}{\tau_{_D}}\bigg)^3 \ \ \ \ OR \ \ \ \ \tan(\theta) = \bigg(\frac{\ell}{\ell_{_D}}\bigg)^{3/2}
\end{equation}

In the thermal limit, the $\delta K$ terms become negligible. 

\begin{figure}[h]
    \includegraphics[scale=0.8]{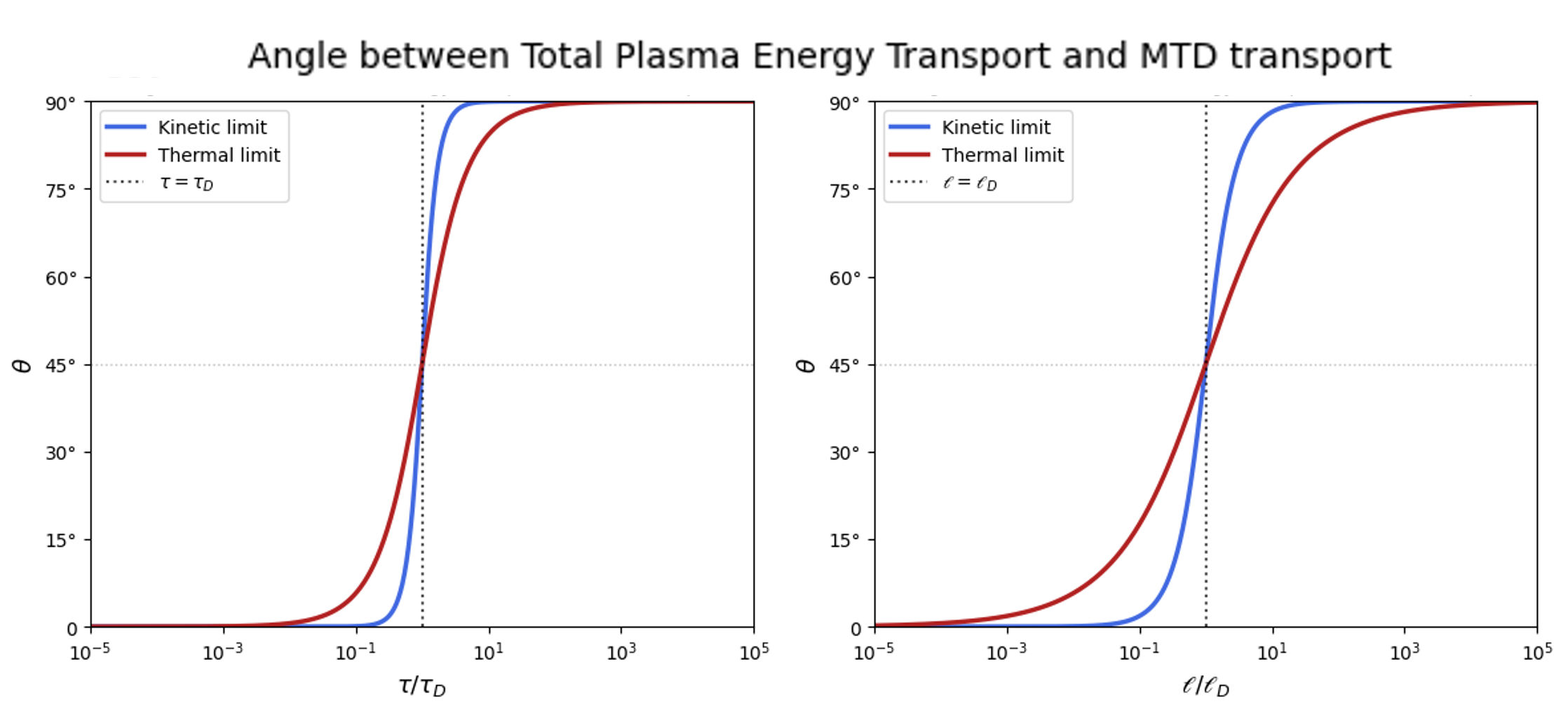}
    \caption{Scale-dependence of plasma energy transport on $\tau_{_D}$ and $\ell_{_D}$.}
    \label{fig:angleplots}
\end{figure}

If $\delta K_{_{E,D}} \ll \delta H_{_{E,D}}$ or $\tau,\ell \ll \tau_{_{KH}},\ell_{_{KH}}$:
\begin{equation} \label{eqn:angleh}
\tan(\theta) = \frac{\delta H_{_E}}{\delta H_{_D}} = \frac{\delta v_{_E}}{\delta v_{_D}}
\end{equation}
\begin{equation}
\tan(\theta) = \frac{\tau}{\tau_{_D}} \ \ \ \ OR \ \ \ \ \tan(\theta) = \sqrt{\frac{\ell}{\ell_{_D}}}
\end{equation}
The dependencies of $\theta$ on the fluctuation and drift scales are shown in figure \ref{fig:angleplots}

It is more difficult to conceptualize $\tau_{_{KS,HS}}$ and $\ell_{_{KS,HS}}$,  but one can examine how they vary under certain limits.  In the limits of $\delta E,n \rightarrow 0$ and $B\rightarrow \infty\;,\;\;  \tau_{_{KS,HS}} \ ,\ \ell_{_{KS,HS}} \rightarrow \infty$ and the magnetic energy transport is much more efficient. Low particle densities limit the amount of carriers available to transport particle energy and weak electric fields limit the transport mechanism itself.  

The maximum power density $\delta J_{_E}\delta E$ can also be determined, allowing for direct comparison with other power densities. For an arbitrary energy flux density $\vec{Q}$,  $\nabla\cdot \vec{Q}$ corresponds to $\delta Q/\ell$ and for an arbitrary energy density $u$, $\partial_t u$ corresponds to $\delta u/\tau$.  With equation \ref{eqn:deltaj} and the energy fluxes in equations \ref{eqn:kterms}-\ref{eqn:driftterms}, $\ell$ and $\tau$ can be determined for a variety of equality conditions:

\begin{table}[h]
\centering
\renewcommand{\arraystretch}{2.5}
\setlength{\tabcolsep}{12pt}
\begin{tabular}{c|c|c}
\hline\hline
\textbf{Condition} & \textbf{Fluctuation Form} & \textbf{Critical Scale} \\
\hline
$|\vec{J}\cdot\vec{E}| = |\nabla\cdot\vec{K}_{_E}|:$ & 
$\delta J_{_E}\delta E = \delta K_{_E}/\ell$ & 
True for all $\ell$ \\[10pt]
\hline
$|\vec{J}\cdot\vec{E}| = |\nabla\cdot\vec{H}_{_E}|:$ & 
$\delta J_{_E}\delta E = \delta H_{_E}/\ell$ & 
$\ell_{_{KH}}$ \\[10pt]
\hline
$|\vec{J}\cdot\vec{E}| = |\nabla\cdot\vec{K}_{_D}|:$ & 
$\delta J_{_E}\delta E = \delta K_{_D}/\ell$ & 
$\ell_{_D}$ \\[10pt]
\hline
$|\vec{J}\cdot\vec{E}| = |\nabla\cdot\vec{H}_{_D}|:$ & 
$\delta J_{_E}\delta E = \delta H_{_D}/\ell$ & 
$\ell_{_{KH}}^{2/3}\ell_{_D}^{1/3}$ \\[10pt]
\hline
$|\vec{J}\cdot\vec{E}| = |\nabla\cdot\vec{S}|:$ & 
$\delta J_{_E}\delta E = \delta S/\ell$ & 
$\ell_{_{KS}}$ \\[10pt]
\hline
$|\vec{J}\cdot\vec{E}| = |\overleftrightarrow{p}\nabla\cdot\vec{v}|:$ & 
$\delta J_{_E}\delta E = p\,\delta v_{_E}/\ell$ & 
$\dfrac{2}{5}\ell_{_{KH}}$ \\[10pt]
\hline
$|\overleftrightarrow{p}\nabla\cdot\vec{v}| = |\nabla\cdot\vec{H}_{_E}|:$ & 
$p\,\delta v_{_E}/\ell = \delta H_{_E}/\ell$ & 
No solution: $p\,\delta v_{_E} = \frac{2}{5}\delta H_{_E}$ for all $\ell$ \\[10pt]
\hline\hline
$|\vec{J}\cdot\vec{E}| = |\partial_t u_{_K}|:$ & 
$\delta J_{_E}\delta E = \delta u_{_K}/\tau$ & 
No solution: $\delta J_{_E}\delta E = 2\delta u_{_K}/\tau$ for all $\tau$ \\[10pt]
\hline
$|\vec{J}\cdot\vec{E}| = |\partial_t u_{_T}|:$ & 
$\delta J_{_E}\delta E = \delta u_{_T}/\tau$ & 
$\sqrt{\frac{3}{10}}\tau_{_{KH}}$ \\[10pt]
\hline
$|\vec{J}\cdot\vec{E}| = |\partial_t u_{_P}|:$ & 
$\delta J_{_E}\delta E = \delta u_{_P}/\tau$ & 
$\sqrt{\frac{3}{5}}\tau_{_{KH}}$ \\[10pt]
\hline
$|\vec{J}\cdot\vec{E}| = |\partial_t u_{_E}|:$ & 
$\delta J_{_E}\delta E = \delta u_{_E}/\tau$ & 
$\tau_{_E} = \dfrac{1}{\sqrt{2}\,\omega_p}$ \\[10pt]
\hline
$|\vec{J}\cdot\vec{E}| = |\partial_t u_{_M}|:$ & 
$\delta J_{_E}\delta E = \delta u_{_M}/\tau$ & 
$\tau_{_M} = \tau_{_E}\dfrac{Bc}{\delta E}$ \\[10pt]
\hline\hline
\end{tabular}
\caption{Critical scales derived from comparisons between the characteristic power density $\delta J_{_E}\delta E$ and energy flux divergences (upper section) or energy density time derivatives (lower section). The middle column expresses each condition in terms of characteristic fluctuation quantities. }
\label{tab:jdotescales}
\end{table}

\begin{figure}[h]
    \includegraphics[scale=0.7]{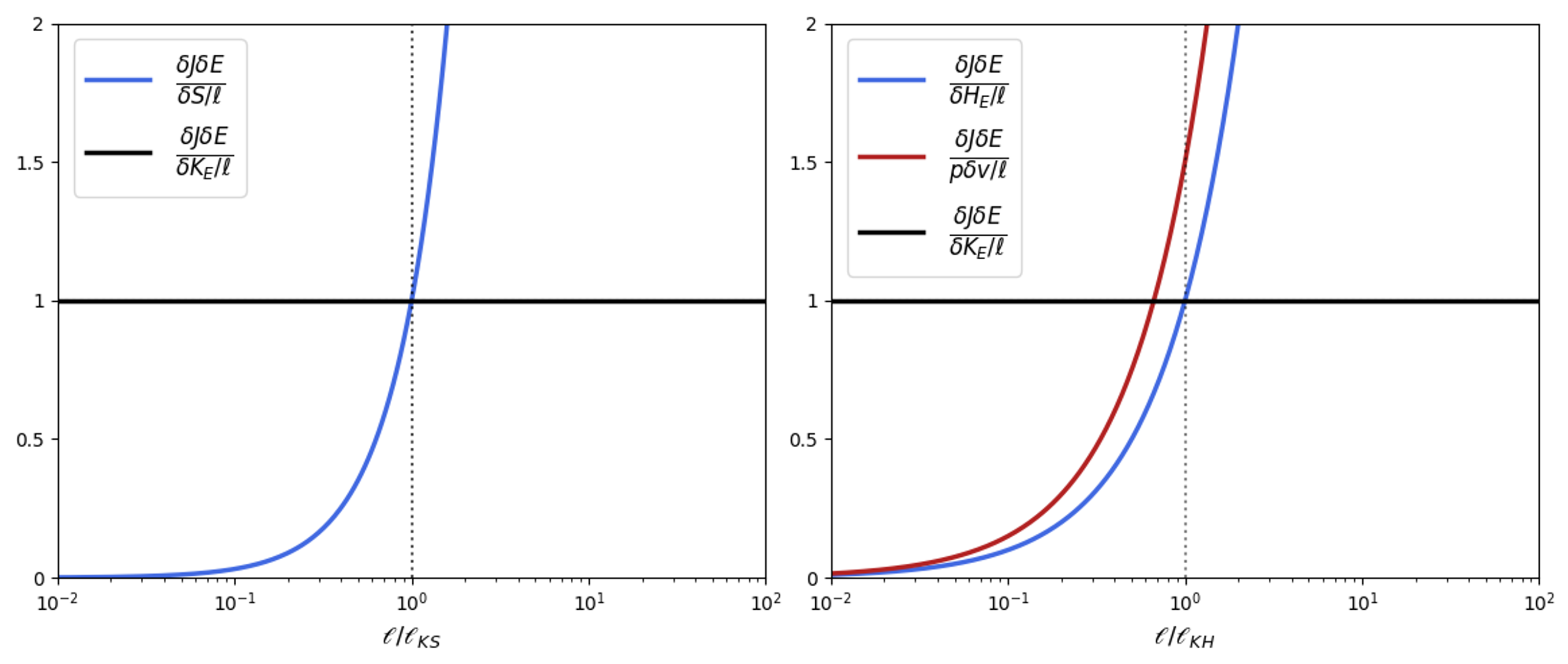}
    \caption{Scale-dependence of power density ratios on $\ell_{_{KS}}$ and $\ell_{_{KH}}$.}
    \label{fig:powerplots}
\end{figure}

The first interesting result in table \ref{tab:jdotescales} is that for \textit{any} $\ell$, $\delta J_{_E}\delta E=\delta K_{_E}/\ell$, reaffirming that $\delta K_{_E}$ is the only term among equations \ref{eqn:kterms}-\ref{eqn:driftterms} associated with direct energy exchange between the electric and plasma subsystems $E\leftrightarrow P$.  There are two other comparisons which reveal scalings independent of $\ell$ or $\tau$.  $p\delta v_{_E} = \frac{2}{5}\delta H_{_E}$ for any $\ell$ and $\delta J_{_E}\delta E = 2u_{_K}/\tau$ for any $\tau$, indicating direct relationships between enthalpy flux divergence and pressure strain and between energy exchange and temporal evolution of bulk energy.  Most of the other conditions in table \ref{tab:jdotescales} can be satisfied at or near one of the critical scales defined in table \ref{tab:criticalscales}. The only condition not satisfied by any individual previously-defined critical lengthscale is $\delta J_{_E}\delta E = \delta H_{_D}/\ell$, which comes out to a geometric mean of the $\ell_{_{KH}}$ and $\ell_{_D}$ length scales.  The plasma frequency $\omega_p$ is directly tied to the new $\tau_{_E}$ critical timescale, which also helps define the new $\tau_{_M}$ critical scale.   Examples of some of the dependencies in table \ref{tab:jdotescales} are plotted in figure \ref{fig:powerplots}.

\section{\label{sec:emergent} Emergent Dynamics}
This section brings together the energy continuity concepts of section \ref{sec:systems} and the scale-dependencies of section \ref{sec:scales} to describe how systems with many small-scale MTD and ETD interactions lead to emergent macroscopic energy transport and transfer.

\subsection{Non-Uniform Closed Systems}

Consider a closed macroscopic system which consists of only plasma particles and stochastic electric field fluctuations due to their own thermal motion. Within this system, local small-scale $\delta E$ fluctuations exchange and transport energy according to the ETD mechanisms described in Section \ref{sec:critscales}.  The macroscopic behavior of the system is determined by the combined effect of all such interactions.  If the $\delta E$ fluctuations are isotropic, the local $\delta \vec{K}_{_E}$, $\delta \vec{H}_{_E}$, and $\delta\vec{J}_{_E}$ contributions cancel on average and the macroscopic ETD transport becomes negligible.  If there exist spatial gradients such as a net $\nabla n \neq 0$, the ETD fluctuations will be larger on the high-density side and result in a macroscopic $\vec{K}_{_E}$ and $\vec{H}_{_E}$ toward the low-density side.  With time, the ETD fluctuations will increase in the low density region and decrease in the high density region as the system asymptotically approaches a state where $\vec{J}\cdot\vec{E}\rightarrow0$ and $\nabla\cdot \vec{K}_{_E}\rightarrow0$,$\nabla\cdot \vec{H}_{_E}\rightarrow0$ everywhere. Although the $\delta K_{_E}$ and $\delta H_{_E}$ fluctuations can be locally significant in the asymptotic limit, their cancellation across large scales ensures no net energy transfer between subsystems occurs. This asymptotic state of the ETD variables will be referred to as a global electro-thermodynamic equilibrium (ETE).

Now consider a magnetic field with a strong gradient $\nabla|B| \perp \vec{B}$ imposed on a system in ETE.  This introduces a non-isotropic distribution of $\delta \vec{E}\times \vec{B}$ and therefore $\delta S$, $\delta K_{_D}$, and $\delta H_{_D}$ fluctuations, resulting in a net macroscopic $\vec{S}$,$\vec{K}_{_D}$, and $\vec{H}_{_D}$ pointing down the gradient in $|B|$. If the $\delta E$ fluctuations are small ($\tau,\ell \ll \tau_{_D},\ell_{_D}$) the MTD mechanisms dominate the ETD mechanisms and this net MTD transport carries plasma and electromagnetic energy toward the region of smaller $|B|$. MTD transport pushes the system \textit{away} from ETE while driving it toward a global magneto-thermodynamic equilibrium (MTE), where $\nabla\cdot \vec{S}\rightarrow0$, $\nabla\cdot \vec{K}_{_D}\rightarrow0$, $\nabla\cdot \vec{H}_{_D}\rightarrow0$, and $\partial_t u_{_{EM}}\rightarrow 0 $ everywhere.  Once the system achieves MTE, the MTD transport mechanisms cancel on average and the weaker ETD mechanisms continue to drive the system back toward ETE on a longer timescale while MTE is globally maintained. The system eventually reaches a state of MTE and ETE, after which all macroscopic energy transport and transfer ceases.

\begin{figure}[h]
    \includegraphics[scale=0.6]{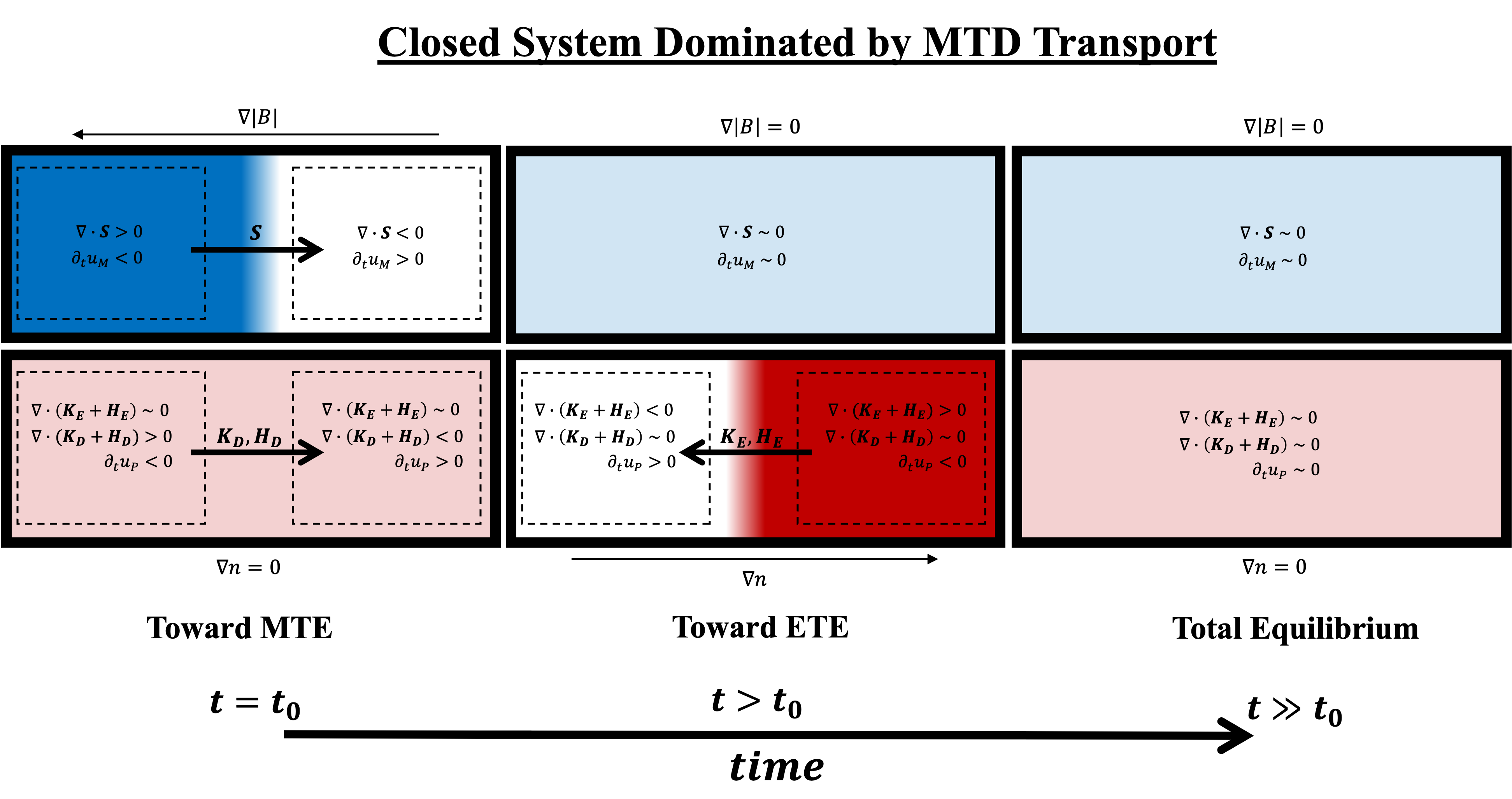}
    \caption{Equilibration of a MTD-dominant system in initial ETE.  The top row shows the evolution of magnetic energy density (shaded blue) over time, and the bottom row shows the evolution of plasma energy density (shaded red).}
   \label{fig:MTE}
\end{figure} 

\subsection{Local Equilibrium}
Most systems of interest are not closed, therefore the concepts of global ETE and MTE may not be particularly useful for the study of localized plasma processes.  Conditions of \textit{local} ETE and MTE can be similarly defined.  
\begin{equation}\label{eqn:localequilibria}
\text{Local Equilibria:} 
\begin{cases}
\vec{J}\cdot\vec{E}\sim\nabla \cdot \vec{K}_{_E} \sim\nabla \cdot \vec{H}_{_E}\sim0 & \text{(ETE) } \\ \\
   \nabla \cdot \vec{S} \sim \nabla \cdot \vec{K}_{_D} \sim \nabla \cdot\vec{H_{_D}}\sim 0    & 
 \text{(MTE) } 
\end{cases}
\end{equation}

\begin{figure}[h]
    \includegraphics[scale=0.6]{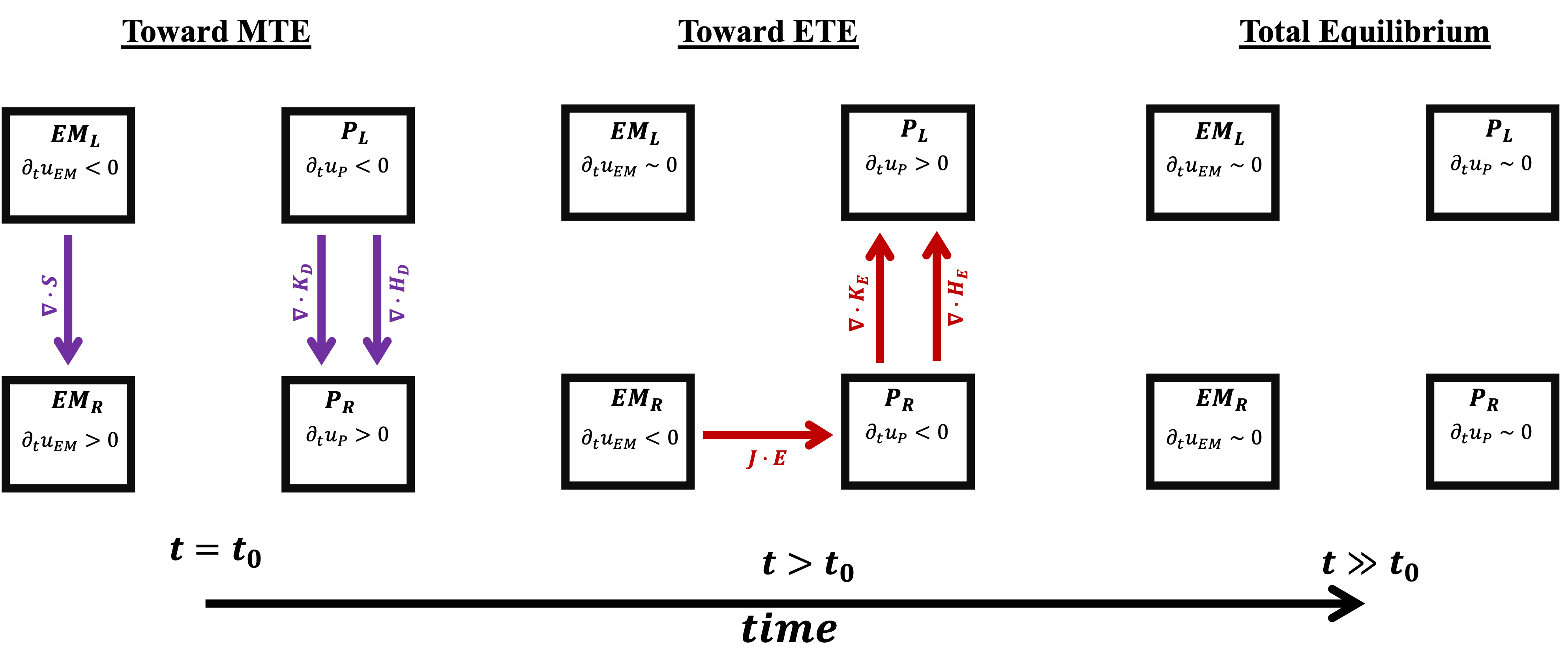}
    \caption{System diagrams depicting energy exchange between the left (L) and right (R) sides of the domain depicted in figure \ref{fig:MTE} as the system equilibrates.  The arrow colors represent ETD (red) or MTD (purple) mechanisms and their direction indicates the direction of energy transport.}
   \label{fig:LocalETEMTE}
\end{figure}

Local ETE and MTE do not require local energy densities to be static.  The system in figure \ref{fig:MTE} is initially in global ETE and subject to MTD mechanisms which eventually drive it out of ETE.  The dashed boxes represent systems initially in local ETE because the divergence of the ETD terms is zero.  However, since the divergence of the MTD terms on each side is nonzero, the plasma energy density in both systems is evolving.  The local systems evolve back toward ETE, converting some energy via net $\vec{J}\cdot\vec{E}$ contributions.  Figure \ref{fig:LocalETEMTE} includes simplified diagrams similar to those in figure \ref{fig:localsystems}, showing how the local systems are linked when approaching ETE vs MTE.

Local ETE and MTE are useful constructs even if the region of interest is in neither state. The fluctuation and critical scales of section \ref{sec:critscales} determine whether ETD or MTD mechanisms dominate the local energy transport. The dominant transport mechanism determines which equilibrium the local system is closest to and therefore how it evolves. This hierarchy is established by the ordering of the critical scales relative to each other and to the characteristic $\delta E$ fluctuation scales.  If $\tau\ll\tau_{_{D,KS}}$,$\ell\ll\ell_{_{D,KS}}$ MTD mechanisms dominate and MTE is the nearest equilibrium.  If $\tau\gg\tau_{_{D,KS}}$,$\ell\gg\ell_{_{D,KS}}$, ETD mechanisms dominate and ETE is the nearest equilibrium.  The smallest critical scale represents the lowest threshold the fluctuation scales must cross, and therefore the first equilibrium the system will naturally prioritize.

\section{\label{sec:reconnection}Magnetic Reconnection}
The preceding sections discussed the energy continuity relationships between systems and subsystems, the local energy fluxes and critical scales that determine which terms dominate energy continuity, and emergent behavior of systems defined by their critical scales, including new types of quasi-equilibria.  In this section, the principles of those previous are invoked to describe the process of magnetic reconnection in an antiparallel field reversal.  For simplicity, the following arguments focus specifically on the evolution of the critical length scales $\ell_{_{KS}}$ and $\ell_{_D}$, and how they relate to the balance between $\nabla\cdot\vec{S}$, $\vec{J}\cdot\vec{E}$, and contributions to $\nabla\cdot\vec{K}$ while ignoring $\nabla\cdot\vec{H}$ contributions.  

\subsection{Pre-Onset}
Consider a system which consists of a  magnetic field reversal embedded within a MTD-dominant reservoir of magnetic and plasma energy upstream.  Since the system is completely in the MTD regime initially, $\vec{J}\cdot\vec{E}$ is negligible in the energy continuity equations  \ref{eqn:endencon1} and \ref{eqn:endencon2} and there is effectively no energy transport between the E and P subsystems.  One consequence of the magnetic field reversal is that there is a null in $|B|$ relative to the upstream at the center, and therefore if an isotropic background $\delta E$ distribution is assumed, there is a non-uniform distribution of $\delta S$, $\delta K_{_D}$, $\delta H_{_D}$ fluctuations which result in net MTD transport that points toward the center of the reversal.    
In the absence of appreciable $\vec{J}\cdot\vec{E}$ and $\vec{K}_{_E}$, the net convergence of $\nabla\cdot\vec{S}$ toward the center of the reversal results in $\partial_tu_{_{EM}} >0$ and the convergence of $\nabla\cdot\vec{K}_{_D}$ results in $\partial_tu_{_P} >0$.  Since $\vec{J}\cdot\vec{E}$ is initially negligible, we can also assume in this case that $T$ is not changing and therefore the evolution of $u_{_P}$ is due to $\partial_t n > 0$.  It is expected that most of the accumulation of electromagnetic energy is an accumulation of magnetic energy density $\partial_t u_{_M}>0$, however, the center of an antiparallel magnetic field reversal represents a topological constraint which prevents further buildup of the magnetic energy in its immediate vicinity.  Thus, if there is any net convergence of  $\nabla\cdot\vec{S}$ very close to the center, $\partial_t u_{_{EM}} \sim \partial_t u_{_E} \propto \partial_t(\delta E^2)>0$ . Since $\ell_{_{KS}}\propto n^{-2/3}\delta E^{-1/3}$ and $\ell_{_D} \propto \delta E$ (table \ref{tab:criticalscales}), the conditions $\partial_t (\delta E) >0$ and $\partial_t n>0$ imply that $\partial_t \ell_{_{KS}}<0$ and $\partial_t \ell_{_D}>0$.  This process therefore causes $\ell_{_{KS}} \rightarrow \ell$ and $\ell_{_D} \rightarrow \infty$.  

\begin{align}
\text{Pre-Onset Case: } \ell&< \ell_{_{KS}}, \ell_{_D}:  \notag \\
|\vec{J}\cdot\vec{E}|,|\nabla\cdot\vec{K}_{_E}|  &< |\nabla\cdot\vec{S}|, |\nabla\cdot\vec{K}_{_D}| \\ \notag \\
\partial_t u_{_E} &=-\vec{J}\cdot\vec{E}- \nabla \cdot \vec{S} \sim - \nabla \cdot \vec{S}>0 \\
\partial_t u_{_P} &=(\vec{J}\cdot\vec{E}-\nabla \cdot \vec{K}_{_E})- \nabla \cdot \vec{K}_{_D} \sim - \nabla \cdot \vec{K}_{_D}>0  \\ 
\notag \\
\partial_t \ell_{_{KS}} &< 0 \Rightarrow \ell \leftarrow \ell_{_{KS}} \\ 
\partial_t \ell_{_{D}} &> 0 \Rightarrow \ell_{_{D}} \rightarrow \infty 
\end{align}

So long as the inflow $K_{_D}$ and $S$ values remain fixed, the critical scales will continue to evolve until $\ell\sim \ell_{_{KS}}$.  

\subsection{Onset}
\begin{figure}[h]
    \centering
    \includegraphics[scale=0.5]{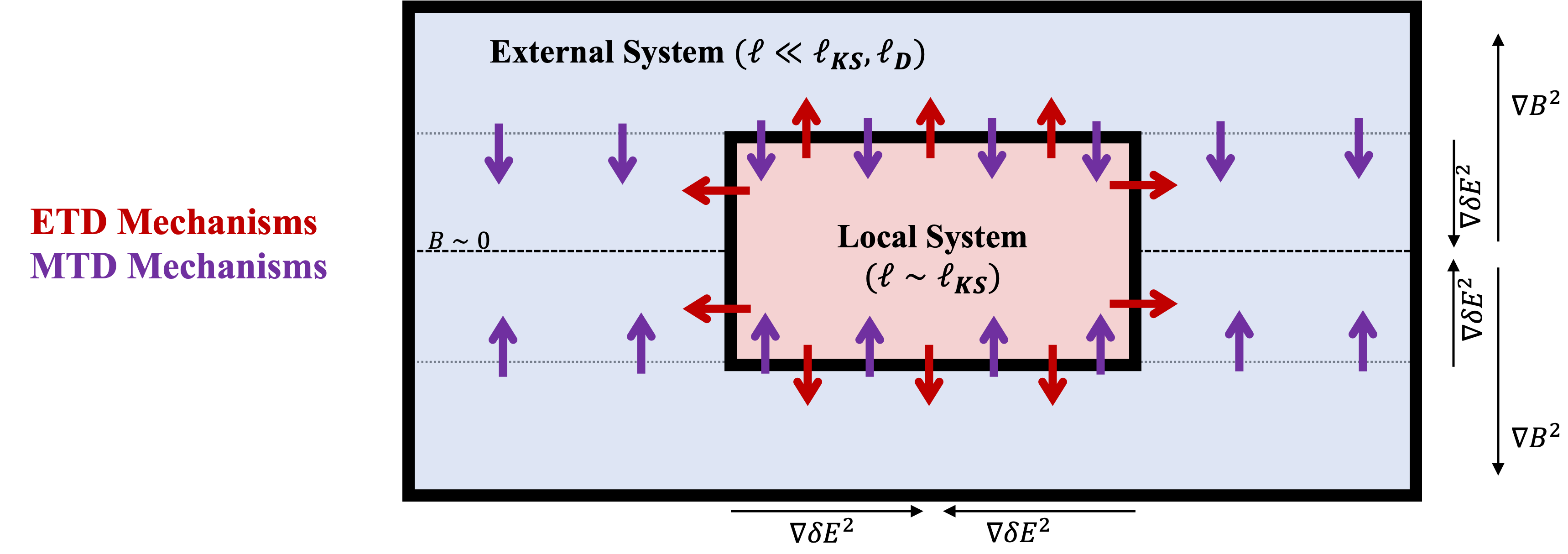} 
    \includegraphics[scale=0.4]{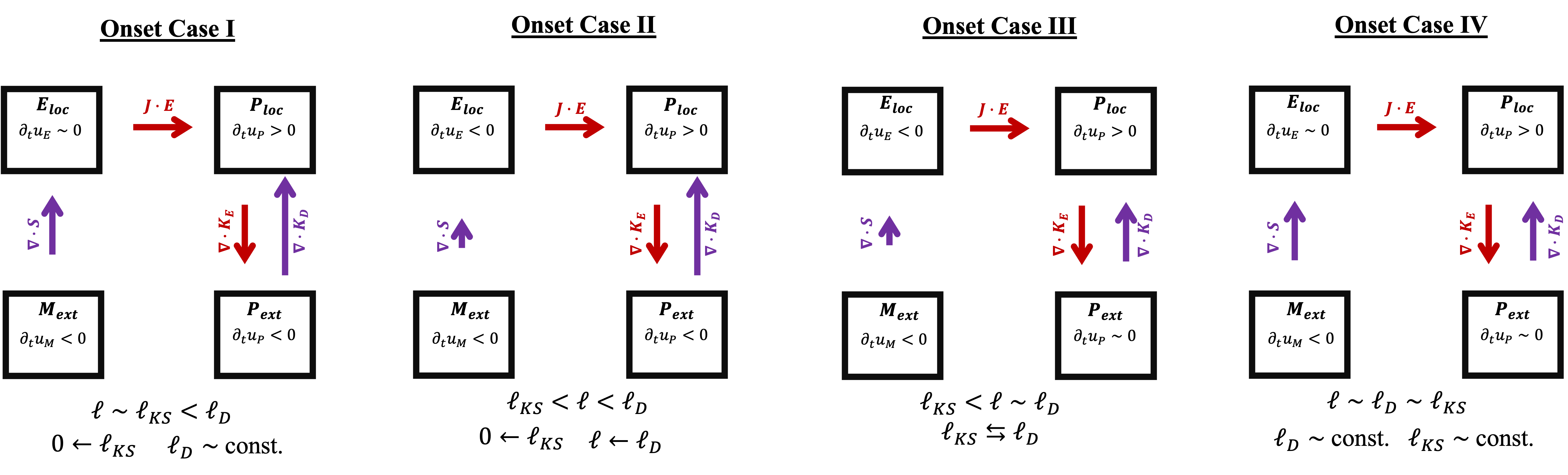} 
    \caption{Structure of the energy transport mechanisms during onset and system diagrams for three onset cases.  The lengths of the arrows depict the magnitude of their respective mechanisms relative to each other.}
    \label{fig:onset}
\end{figure}

The condition $\ell\sim\ell_{_{KS}}$ marks a transition in the dynamics of local energy transport.  The ETD transport of kinetic energy away from the electric field compression region becomes comparable in magnitude to the MTD transport of electromagnetic energy from the inflows.  From table \ref{tab:jdotescales}, this also implies that $|\vec{J}\cdot\vec{E}|\sim|\nabla\cdot\vec{S}|$, and therefore the electric field energy neither accumulates nor depletes $\partial_t u_{_E}\sim 0$.  If $\ell_{_D}>\ell$,  the density is still accumulating and therefore $\ell_{_{KS}}$ will continue to decrease. 
\begin{align}
\text{Onset Case I: } \ell &\sim\ell_{_{KS}} < \ell_{_D}:  \notag \\
|\vec{J}\cdot\vec{E}|,|\nabla\cdot\vec{K}_{_E}|&\sim  |\nabla\cdot\vec{S}|< |\nabla\cdot\vec{K}_{_D}| \\ \notag \\
\partial_t u_{_E} &=-\vec{J}\cdot\vec{E}- \nabla \cdot \vec{S} \sim 0 \\
\partial_t u_{_P} &\sim- \nabla \cdot \vec{K}_{_D} >0 \\ 
\notag \\
\partial_t \ell_{_{KS}} &< 0 \Rightarrow 0 \leftarrow \ell_{_{KS}} \\ 
\partial_t \ell_{_{D}} &\sim 0 \Rightarrow \ell_{_{D}} \sim \text{const.} 
\end{align}
After some $\Delta t$, the hierarchy of critical scales becomes $\ell_{_{KS}}<\ell<\ell_{_D}$.  Now the dissipation of energy and its transport out of the local system by $\vec{K}_{_E}$ exceeds the transport of electromagnetic energy into the local system by $\vec{S}$, leading to a depleting electric field energy $\partial_t u_{_E} < 0$, which causes $\ell_{_D}$ to decrease and approach $\ell$.  Since $\ell_{_D}>\ell_{_{KS}}$ and $\ell_{_D}>\ell$, the increase in $n$ dominates the evolution of $\ell_{_{KS}}$, which continues to decrease away from $\ell$.   
\begin{align}
\text{Onset Case II: } \ell_{_{KS}} &<\ell < \ell_{_D}:  \notag \\
|\nabla\cdot\vec{S}|&<|\vec{J}\cdot\vec{E}|,|\nabla\cdot\vec{K}_{_E}|< |\nabla\cdot\vec{K}_{_D}| \\ \notag \\
\partial_t u_{_E} &\sim-\vec{J}\cdot\vec{E} < 0 \\
\partial_t u_{_P} &\sim- \nabla \cdot \vec{K}_{_D} >0 \\ 
\notag \\
\partial_t \ell_{_{KS}} &< 0 \Rightarrow 0 \leftarrow \ell_{_{KS}} \\ 
\partial_t \ell_{_{D}} &< 0 \Rightarrow \ell \leftarrow \ell_{_D}
\end{align}
The next state of the critical scales after sufficient $\Delta t$ then becomes $\ell_{_{KS}}<\ell\sim\ell_{_D}$.  Since $\ell\sim\ell_{_D}$, the density is no longer evolving and therefore $\ell_{_{KS}}$ is increasing due to the decrease in $\delta E$.  For the same reason, $\ell_{_D}$ decreases.    
\begin{align}
\text{Onset Case III: } \ell_{_{KS}}&<\ell \sim \ell_{_D}:  \notag \\
|\nabla\cdot\vec{S}|&<|\vec{J}\cdot\vec{E}|,|\nabla\cdot\vec{K}_{_E}|\sim |\nabla\cdot\vec{K}_{_D}| \\ \notag \\
\partial_t u_{_E} &\sim-\vec{J}\cdot\vec{E}< 0 \\
\partial_t u_{_P} &\sim(\vec{J}\cdot\vec{E}-\nabla \cdot \vec{K}_{_E})- \nabla \cdot \vec{K}_{_D} \sim \vec{J}\cdot\vec{E} > 0 \\ 
\notag \\
\partial_t \ell_{_{KS}} &> 0 \Rightarrow \ell_{_{KS}} \rightarrow \ell_{_D}  \\ 
\partial_t \ell_{_{D}} &< 0 \Rightarrow \ell_{_{KS}} \leftarrow \ell_{_{D}} 
\end{align}
Now the local system is evolving such that the $\ell_{_{KS}}$ and $\ell_{_D}$ critical scales approach each other. The diagram in figure \ref{fig:onset} illustrates the configuration of the system and the onset cases.  Onset cases I-III collectively describe a system which is evolving toward an ever-changing equilibrium, cycling trough different cases over time.  The combined effect is that, without any variation in the inflow parameters, the system asymptotically approaches a state where $\ell_{_D}\sim\ell\sim\ell_{_{KS}}$.  In this configuration, the density and electric field energy are not evolving and therefore $\ell_{_D}$ and $\ell_{_{KS}}$  are not evolving.  However, the non-negligible $\vec{J}\cdot\vec{E}$ still contributes to an increase in the plasma energy density.
\begin{align}
\text{Onset Case IV: } \ell_{_{D}}&\sim\ell \sim \ell_{_{KS}}:  \notag \\
|\nabla\cdot\vec{K}_{_D}|&\sim|\vec{J}\cdot\vec{E}|,|\nabla\cdot\vec{K}_{_E}|\sim|\nabla\cdot\vec{S}|  \\ \notag \\
\partial_t u_{_E} &\sim-\vec{J}\cdot\vec{E}- \nabla \cdot \vec{S} \sim 0 \\
\partial_t u_{_P} &\sim(\vec{J}\cdot\vec{E}-\nabla \cdot \vec{K}_{_E})- \nabla \cdot \vec{K}_{_D} \sim \vec{J}\cdot\vec{E} > 0 \\ 
\notag \\
\partial_t \ell_{_{KS}} &\sim 0 \Rightarrow \ell_{_{KS}}\sim \text{const.}  \\ 
\partial_t \ell_{_{D}} &\sim 0 \Rightarrow  \ell_{_{D}}\sim \text{const.} 
\end{align}

As long as the MTD inflow is held constant, the critical scales continue to evolve with respect to one another and with respect to the characteristic scale $\ell$ of the ETD interactions.  The cases above are not exhaustive (see appendix for the onset cases starting from the condition $\ell_{_D}<\ell\sim\ell_{_{KS}}$), but they all tend to approach a dynamic equilibrium where the critical scales and characteristic scales are comparable, and where external magnetic energy is efficiently transferred to local plasma energy.  This process can be understood as a consequence of incompatible equilibria.  The local system is dominated by ETD transport, and therefore the equilibrium toward which it is evolving is ETE.  However, it is interacting with an external system which is MTD-dominant and therefore is moving toward MTE.  As the external system loses energy via MTD transport, the lost energy maintains the local $u_{_E}$ and net ETD transport out of the local system.  Thus, the local system is always moving toward ETE without ever reaching it, so long as there is a sufficient reservoir of MTD-dominated plasma in the external system which can maintain constant inflow.  

\section{\label{sec:discussion} Discussion}
Sections \ref{sec:systems}-\ref{sec:reconnection} presented a framework for describing the transport and transfer of energy in collisionless plasmas.  The four main concepts are summarized below.
\begin{description}
    \item[Systems and Subsystems] Electromagnetic fields and plasmas are treated as separate interacting subsystems through which energy can flow, accumulate, or deplete.  Disconnected subsystems can only exchange energy through another subsystem ($M\leftrightarrow E\leftrightarrow K\leftrightarrow T$).
    \item[Scale-Dependent Energy Transport] The relative influence of connections between systems is dependent on the characteristic scales of the electrostatic field fluctuations, which contribute to competing ETD and MTD energy transport mechanisms.  Critical scales determine the threshold beyond which one mechanism exceeds another.  
    \item[Emergent Dynamics] Stochastic ETD and MTD transport tends to smooth energy density gradients with time.  If there are opposing gradients, then the system's critical scales determine which transport mechanisms dominate and therefore what type of equilibrium is prioritized. 
    \item[Magnetic Reconnection] Dynamic equilibrium can occur by coupling a local system seeking ETE to an external system seeking MTE,  allowing MTD transport to be efficiently converted to local ETD transport.     
\end{description}

One of the central components on which everything else depends is the nature of the stochastic electric field and, most importantly, its spatiotemporal scales.  While many of the critical scales in table \ref{tab:criticalscales} are not obvious without working through the derivation, the $\tau_{_D}$ scale identifies the more intuitive connection between cyclotron frequency and the nature of plasma energy transport.  It has recently been shown \cite{mallet2026perpendicular} that a similar term can be used to determine the condition of ion heating/acceleration via magnetic moment breaking.  The importance of electrostatic field structures to the energy dissipation mechanisms of reconnection and turbulence has also been well-demonstrated \cite{ergun2020particle,vo2026kinetic}.  The other critical scales in table \ref{tab:criticalscales} were derived via similar procedures as the $\tau_{_D}$ scale, but represent a variety of different energy flux comparisons.

The work presented  has not addressed the specific dynamics of kinetic scale behavior, higher order moments, heat flux, or kinetic entropy, which all require a more detailed analysis from a modern kinetic theory approach.  The scale dependencies derived in section \ref{sec:scales} do have a kinetic origin, but there is no consideration of individual particle dynamics or phase space evolution in the arguments of sections \ref{sec:emergent} and \ref{sec:reconnection}.  Energy is the primary agent which moves through the system and subsystems via fluctuations in electromagnetic fields.    The characteristic energy channels of any given system of interest are determined by its average critical scales.  The unique dissipative structures of collisionless plasmas are often multiscale, therefore it is useful to consider the spatial distribution of critical scales.  In principle, the distribution of critical scales in a given system determines the future evolution of energy throughout the system and subsystems at all spatiotemporal scales.  Predictions based on this principle are inherently limited in practice due to imperfect knowledge of the system and computational resource constraints, but it may still be useful for understanding multiscale mechanisms of energy dissipation.  Section \ref{sec:reconnection} showed that under a few simple assumptions, the eventual condition of reconnection onset and converging of the critical scales $\ell_{_{KS}}$ and $\ell_{_D}$ within an ETD-dominant diffusion region are predicted.  The arguments of section \ref{sec:reconnection} did not distinguish separate particle species, and therefore did not distinguish the IDR and EDR.  A multi-species treatment would require multiple sets of critical scales and additional subdivisions of the plasma system for each species and therefore complicate the analysis.  

By ignoring enthalpy flux, pressure-strain, and the critical scales associated with them, section \ref{sec:reconnection} described a special case of reconnection onset in the kinetic limit, defined by the condition $\ell\gg\ell_{_{KH}}$.  However, the terms associated with thermal energy transport and conversion are important to consider in pursuit of a more comprehensive description of reconnection.  As $T$ is increased, $\ell_{_{KH}}$ increases linearly with it according to table \ref{tab:criticalscales}.  A sufficiently large $T$ can therefore put the system in the thermal limit ($\ell\ll\ell_{_{KH}}$), which also implies that $|\overleftrightarrow{p}\cdot\nabla\vec{v}_{_E}|\gg|\vec{J}\cdot\vec{E}|$ according to table \ref{tab:jdotescales}.  A large increase in $T$ also yields a nonlinear ($\propto T^{-2}$) decrease in $\ell_{_{HS}}$, dramatically reducing the threshold necessary for $\delta\vec{H}_{_E}$ terms to dominate $\delta \vec{S}$ terms.  The thermal onset condition directly analogous to the kinetic onset condition of section \ref{sec:reconnection} is then $\ell\sim\ell_{_{HS}}$, meaning that $|\nabla\cdot\vec{S}|\sim |\nabla\cdot \vec{H}_{_E}|$ with negligible (but still nonzero) contributions from $|\vec{J}\cdot\vec{E}|$ and $|\nabla\cdot\vec{K}_{_E}|$.   A local system balanced in such a way may still take on a similar geometry as in figure \ref{fig:onset}, where the characteristic local energy transfer is between bulk kinetic and thermal energy ($K\rightarrow T$) rather than between electric and kinetic energy ($E\rightarrow K$).   However, the electric field is still the necessary mechanism by which thermal energy is evacuated, so it still must be sustained by $\nabla \cdot \vec{S}$ for the process to endure.  It is unclear how a system in the thermal limit is sustainable if $|\overleftrightarrow{p}\cdot\nabla\vec{v}_{_E}|\gg|\vec{J}\cdot\vec{E}|$ and the bulk kinetic energy is depleted faster than acceleration by electric fields can replenish it.  It may be the case that the mutual influence of the evolving critical scales due to energy flux imbalances also pushes the system toward a dynamic equilibrium similar to that  of section \ref{sec:reconnection}.   The difference between a reconnection site in the thermal vs kinetic limit may just be a matter of scale in some cases.  It is clear from figure \ref{fig:angleplots} that the transition from the MTD to ETD regime is more broad in scale in the thermal limit, indicating a more forgiving criteria for ETD mechanisms to begin to manifest. Diffusion regions may exhibit a structure like the one in section \ref{sec:reconnection}, while at much larger scales the reconnection energy budget is dominated by thermal energy since the bulk outflows coming from the diffusion region quickly thermalize during their remagnetization \cite{payne2021origin}.  Studies of turbulent dissipation have also established that signatures of large  $|\overleftrightarrow{p}\cdot\nabla\vec{v}_{_E}|$ and $|\vec{J}\cdot\vec{E}|$ tend to be regionally correlated with local, scale-dependent intermittency regarding which term is the most active 'dissipation' measure\cite{yang2019scale}.  A detailed analysis of the dynamic relationship between the thermal and kinetic limits and their implications for reconnection onset and structure ought to be considered for future theoretical development.

The influence of magnetic geometry on the transport of energy was invoked in section \ref{sec:reconnection} to describe an antiparallel field reversal, but there are a variety of reconnecting structures which vary in magnetic orientation and pressure balance.  A guide field contributes $u_{_M}\neq 0$ and $\delta \vec{E}\times\vec{B}\neq 0$ at the center of a field reversal and can therefore maintain a large $\ell_{_{KS}}$ and small $\ell_{_D}$, suppressing the reconnection onset mechanism of section \ref{sec:reconnection}.  Along with the ratio of thermal to magnetic pressure (plasma beta $\beta$), a finite guide field has indeed been shown \cite{swisdak2003diamagnetic,yoon2024enablement,pyakurel2026dependence} to play a role in the suppression of reconnection onset via diamagnetic suppression or prevention of current sheet collapse.  Despite this, it is clear that reconnection can and does occur under a variety of guide field conditions \cite{dahlin2022variability}.  Kinetic signatures like agyrotropy \cite{swisdak2016quantifying} are suppressed by a guide field while parallel electric fields tend to play a more substantial role in energy dissipation \cite{wilder2018role,nakamura2025outstanding}.  A description of guide field reconnection similar to the analysis of section \ref{sec:reconnection} may therefore require a more careful treatment which takes into account the MTD transport due to the guide field component and the relevance of the enthalpy flux and pressure-strain terms.  

For the sake of brevity, any analysis of critical scales beyond section \ref{sec:critscales} only included the critical length scales and ignored the critical timescales.  The critical length scales relate the energy dissipation terms to energy transport in space, while the critical timescales relate the dissipation terms to the local energy evolution in time.  Clearly, a more robust model of reconnection or turbulence will require a firm understanding of the critical timescales, their implications, and how they tie in to a full spatiotemporal description of collisionless energy dissipation.  For now, this is left as a topic for future exploration.  

One elusive aspect of collisionless energy dissipation is the nature of entropy and its evolution.  The arguments regarding emergent behavior in section \ref{sec:emergent} and the reconnection timeline in section \ref{sec:reconnection} never invoke entropy or any similar parameter to be maximized in some equilibrium state as $t\rightarrow\infty$.  Of course, the very assumption of a preferential arrow of time implies that such a parameter exists in principle, though it is not obvious how one would define it in this context.  A detailed treatment of the spatiotemporal structure of energy dissipation from large to small scales, from electromagnetic energy to kinetic energy, and from kinetic energy to thermal energy has been previously conducted \cite{matthaeus2020pathways}, wherein they argue that the statistical-mechanical Boltzmann entropy and the macroscopic Clausius entropy are decoupled quantities in the absence of collisions so long at the system is not already in local thermodynamic equilibrium (LTE).  Kinetic treatments of collisionless entropy often separate velocity-space and position-space kinetic entropy \cite{liang2020kinetic,argall2022theory} or account for energy conversion in higher-order moments \cite{cassak2023quantifying} for systems far from LTE, though most of these constitute detailed descriptions of the entropy of particle ensembles without additional consideration of the entropy associated with the electromagnetic fields that produce such unique phase-space distributions in the first place.  In the context of the present work, it would be difficult to define a version of entropy which accounts for electromagnetic field evolution and the various types of equilibria presented in section \ref{sec:emergent}, though it may be worthy pursuit for future theoretical work.

\section{\label{sec:conclusion} Conclusion}
This study presents a theoretical framework that attempts to describe collisionless energy transport in a generalized way.  First, by categorizing the different forms energy can take as a network of connected subsystems, it is easier to visualize energy continuity constraints.  Second, by considering the characteristic spatiotemporal scales of a system's fluctuating electric fields, different mechanisms of energy transport can be quantified, and critical scales can be derived which dictate the temporal evolution of energy in a system.  Third, the mechanisms are scaled up to describe the macroscopic transport of electromagnetic and plasma energy driving the system toward preferred equilibria over time.  Finally, the concepts are brought together to describe the formation of a diffusion region in a simplified model of reconnection.  While many simplifications were made for the sake of brevity and clarity, the concepts presented here will hopefully serve as a starting point from which more detailed analyses of collisionless dissipative processes can be conducted.  

\section*{AI Disclosure}
The large language model (LLM) Claude was used to speed up specific, otherwise tedious tasks.  These include proper LaTeX formatting of the tables and writing out of derivations (which were all initially done by hand) in the appendix.  Claude also provided suggestions for altering the ordering of arguments in the manuscript, some of which were incorporated.  All of the actual writing in the body of the manuscript is my own, as are the arguments they express (the earliest versions of this work were briefly discussed in 2021 \cite{payne2021multiscale} and 2023 \cite{payne2023magnetothermodynamic}).

\section*{Appendix}
\appendix
\setcounter{section}{1}
\setcounter{figure}{0}
\subsection{Derivations of Terms in Table \ref{tab:criticalscales}}
Starting with $\delta K_{_E}$ and $\delta H_{_E}$, set them equal and solve for $\tau,\ell$:

\begin{align*}
\text{If }\delta K_{_E} &= \delta H_{_E}: \\ \\
\frac{nm\delta v^3_{_E}}{2} &=  \frac{5nk_bT \delta v_{_E}}{2} \\
m\delta v^2_{_E} &= 2q\delta E \ell = 5k_bT  \\
\ell &= \frac{5k_bT}{2q\delta E} \equiv \ell_{_{KH}} 
\end{align*}
\begin{align*}
m\delta v^2_{_E} = \frac{mq^2\delta E^2 \tau^2}{m^2} &= 5k_bT  \\
\tau^2 &= \frac{5mk_bT}{q^2\delta E^2}  \\
\tau &= \sqrt{\frac{5mk_bT}{q^2\delta E^2}}\equiv \tau_{_{KH}} 
\end{align*}

If $\delta K_{_E} = \delta S$:
\begin{align*}
\sqrt{\frac{2n^2q^3\delta E^3}{m}}\ell^{3/2} &= \frac{\delta E B}{\mu_0} \\
\ell^{3/2} &= \frac{\delta E B}{\mu_0}\sqrt{\frac{m}{2n^2q^3\delta E^3}} \\
&= \frac{B}{\mu_0}\sqrt{\frac{m}{2n^2q^3\delta E}} \\
\ell &= \left(\frac{B}{\mu_0}\right)^{2/3}\left(\frac{m}{2n^2q^3\delta E}\right)^{1/3} \\
&= q^{-1}\left(\frac{B}{\mu_0 n}\right)^{2/3}\left(\frac{m}{2\delta E}\right)^{1/3} \equiv \ell_{_{KS}}
\end{align*}
\begin{align*}
\frac{nq^3\delta E^3}{2m^2}\tau^3 &= \frac{\delta E B}{\mu_0} \\
\tau^3 &= \frac{\delta E B}{\mu_0}\frac{2m^2}{nq^3\delta E^3} \\
&= \frac{2m^2 B}{\mu_0 nq^3\delta E^2} \\
\tau &= \left(\frac{2m^2 B}{\mu_0 nq^3\delta E^2}\right)^{1/3} \\
&= q^{-1}\left(\frac{2B}{\mu_0 n}\right)^{1/3}\left(\frac{m}{\delta E}\right)^{2/3} \equiv \tau_{_{KS}}
\end{align*}

If $\delta K_{_E} = \delta K_{_D}$:
\begin{align*}
\frac{nm\delta v^3_{_E}}{2} &= \frac{nm\delta v^3_{_D}}{2} \\
\delta v_{_E} &= \delta v_{_D} \\
\sqrt{\frac{2q \delta E \ell}{m}} &= \frac{\delta E}{B} \\
\ell &= \frac{\delta E^2}{B^2}\frac{m}{2q\delta E} \\
&= \frac{m\delta E}{2qB^2} \equiv \ell_{_D}
\end{align*}
\begin{align*}
\delta v_{_E} &= \delta v_{_D} \\
\frac{q \delta E \tau}{m}&= \frac{\delta E}{B} \\
\tau &= \frac{m}{qB} \equiv \tau_{_D}
\end{align*}

Similarly for $\delta H_{_E} = \delta H_{_D}$:
\begin{align*}
\frac{5nk_bT\delta v_{_E}}{2} &= \frac{5nk_bT\delta v_{_D}}{2} \\
\delta v_{_E} &= \delta v_{_D} \\ \Rightarrow\ell &= \ell_{_D} \text{ and } \tau = \tau_{_D}
\end{align*}

If $\delta H_{_E} = \delta S$:
\begin{align*}
5nk_bT\sqrt{\frac{q\delta E\ell}{2m}} &= \frac{\delta E B}{\mu_0} \\
\frac{q\delta E\ell}{2m} &= \frac{\delta E^2 B^2}{\mu_0^2(5nk_bT)^2} \\
\ell &= \frac{2m\delta E B^2}{\mu_0^2 q(5nk_bT)^2} \\
&= q^{-1}(2m\delta E)\left(\frac{B}{5\mu_0 nk_bT}\right)^2 \equiv \ell_{_{HS}}
\end{align*}
\begin{align*}
\frac{5nqk_bT\delta E}{2m}\tau &= \frac{\delta E B}{\mu_0} \\
\tau &= \frac{\delta E B}{\mu_0}\frac{2m}{5nqk_bT\delta E} \\
&= q^{-1}\left(\frac{2mB}{5\mu_0 nk_bT}\right) \equiv \tau_{_{HS}}
\end{align*}

\subsection{Derivations of Terms in Table \ref{tab:jdotescales}}

If $\delta J_{_E}\delta E = \delta K_{_E}/\ell$:
\begin{align*}
nq\delta v_{_E}\delta E &= \frac{nm}{2 \ell}\delta v_{_E}^3 \\
q \delta E &= \frac{m}{2\ell} \delta v^2_{_E} = \frac{m}{2\ell}\frac{2q \delta E \ell}{m} \\
q \delta E &= q \delta E \Rightarrow \text{True for any $\ell$} 
\end{align*}

If $\delta J_{_E}\delta E = \delta H_{_E}/\ell$:
\begin{align*}
nq\delta v_{_E}\delta E &= \frac{5nk_bT}{2\ell}\delta v_{_E} \\
q \delta E &= \frac{5k_bT}{2\ell}\\
\ell &= \frac{5k_bT}{2q\delta E} = \ell_{_{KH}}
\end{align*}

If $\delta J_{_E}\delta E = \delta K_{_D}/\ell$:
\begin{align*}
nq\delta v_{_E}\delta E &= \frac{nm}{2\ell}\delta v^3_{_D} \\
q \sqrt{\frac{2q\delta E \ell}{m}}\delta E &= \frac{m}{2\ell} \frac{\delta E^3}{B^3} \\
\ell^{3/2} &= \frac{m}{2} \frac{\delta E^2}{B^3} \sqrt{\frac{m}{2q^3\delta E}} = \frac{1}{B^3} \bigg(\frac{m\delta E}{2q}\bigg)^{3/2} \\
\ell &= \frac{m\delta E}{2qB^2} = \ell_{_D}
\end{align*}

If $\delta J_{_E}\delta E = \delta H_{_D}/\ell$:
\begin{align*}
nq\delta v_{_E}\delta E &= \frac{5nk_bT}{2\ell}\delta v_{_D} \\
q\sqrt{\frac{2q\delta E \ell}{m}}\delta E &= \frac{5k_bT}{2\ell}\frac{\delta E}{B}\\
\sqrt{\frac{2q^3\delta E \ell}{m}}&= \frac{5k_bT}{2B\ell}\\
\ell^{3/2}&= \frac{5k_bT}{2B}\bigg( \frac{m}{2q^3\delta E}\bigg)^{1/2}\\
\ell&= \bigg(\frac{5k_bT}{2B}\bigg)^{2/3}\bigg( \frac{m}{2q^3\delta E}\bigg)^{1/3}\\ \\
\ell_{_{KH}} &= \frac{5k_bT}{2q \delta E} \Rightarrow \frac{5k_bT}{2} = q\delta E\ell_{_{KH}} \\ \\
\ell &= \bigg(\frac{q\delta E\ell_{_{KH}}}{B}\bigg)^{2/3}\bigg( \frac{m}{2q^3\delta E}\bigg)^{1/3} \\
&= \ell_{_{KH}}^{2/3} \bigg( \frac{mq^2 \delta E^2}{2q^3\delta E B^2}\bigg)^{1/3} \\
&= \ell_{_{KH}}^{2/3} \bigg( \frac{m \delta E}{2q B^2}\bigg)^{1/3} = \ell_{_{KH}}^{2/3}\ell_{_D}^{1/3}
\end{align*}

If $\delta J_{_E}\delta E = \delta S/\ell$:
\begin{align*}
nq\delta v_{_E}\delta E &= \frac{\delta E B}{\mu_0 \ell} \\
nq\delta v_{_E} &= \frac{ B}{\mu_0 \ell} \\
nq\sqrt{\frac{2q\delta E \ell}{m}} &= \frac{ B}{\mu_0 \ell} \\
\ell^{3/2} &= \bigg(\frac{mB^2}{2\mu^2_0n^2q^3\delta E}\bigg)^{1/2} \\
\ell&= \bigg(\frac{mB^2}{2\mu^2_0n^2q^3\delta E}\bigg)^{1/3} = q^{-1}\bigg(\frac{B}{\mu_0 n} \bigg)^{2/3} \bigg(\frac{m}{2\delta E} \bigg)^{1/3} = \ell_{_{KS}}\\
\end{align*}

If $\delta J_{_E}\delta E = p\delta v_{_E}/\ell$:
\begin{align*}
nq\delta v_{_E}\delta E &= nk_bT\delta v_{_E}/\ell \\
q\delta E &= k_bT/\ell \\
\ell &= \frac{k_bT}{q \delta E} = \frac{2}{5}\ell_{_{KH}}
\end{align*}

$p\delta v_{_E}/\ell$ vs $\delta H_{_E}/\ell$:

\begin{align*}
p &= nk_bT \\
\frac{\delta H_{_E}}{\ell} &= \frac{5nk_bT}{2\ell}\delta v_{_E} = \frac{5p\delta v_{_E}}{2\ell} \\
\frac{2}{5}\frac{\delta H_{_E}}{\ell} &= \frac{p\delta v_{_E}}{\ell} \Rightarrow \text{True for any $\ell$}
\end{align*}

$\delta J_{_E}\delta E$ vs $\delta u_{_K}/\tau$:
\begin{align*}
\frac{\delta u_{_K}}{\tau} &= \frac{nm\delta v^2_{_E}}{2\tau} = \frac{nq^2\delta E^2\tau}{2m} \\
\delta J_{_E}\delta E&= \frac{nq^2\delta E^2 \tau}{m} = \frac{2\delta u_{_K}}{\tau} \Rightarrow \text{True for any $\tau$}\\
\\
\end{align*}

If $\delta J_{_E}\delta E  =\delta u_{_T}/\tau$:
\begin{align*}
\frac{nq^2\delta E^2 \tau}{m} &= \frac{3nk_bT}{2\tau} \\
\tau &=\bigg( \frac{3mk_bT}{2q^2\delta E^2}\bigg)^{1/2} \\ \\
\tau_{_{KH}} &= \bigg( \frac{5mk_bT}{q^2\delta E^2}\bigg)^{1/2} \\ \\
\Rightarrow \tau &= \sqrt{\frac{3}{10}}\tau_{_{KH}}
\end{align*}

$\delta J_{_E}\delta E$ vs $\delta u_{_P}/\tau$:
\begin{align*}
\frac{nq^2\delta E^2 \tau}{m} &= \frac{nq^2\delta E^2\tau}{2m} +\frac{3nk_bT}{2\tau} \\
\frac{q^2\delta E^2 }{m} &= \frac{q^2\delta E^2}{2m} +\frac{3k_bT}{2\tau^2} \\ 
\frac{q^2\delta E^2 }{2m} &= \frac{3k_bT}{2\tau^2} \\ 
\tau^2 &= \frac{3mk_bT}{q^2\delta E^2} \\ 
\tau &= \bigg(\frac{3mk_bT}{q^2\delta E^2} \bigg)^{1/2} = \sqrt{\frac{3}{5}} \tau_{_{KH}}
\end{align*}

$\delta J_{_E}\delta E$ vs $\delta u_{_E}/\tau$:
\begin{align*}
\frac{nq^2\delta E^2 \tau}{m} &= \frac{\epsilon_0 \delta E^2}{2\tau} \\
\tau^2 &= \frac{\epsilon_0 m }{2nq^2} \\
\tau &= \sqrt{\frac{\epsilon_0 m }{2nq^2}} \\ \\
\omega_p &= \sqrt{\frac{nq^2}{\epsilon_0m}} \\ \\
\Rightarrow \tau &= \frac{1}{\sqrt{2}\omega_p} \equiv \tau_{_E}
\end{align*}

$\delta J_{_E}\delta E$ vs $\delta u_{_M}/\tau$:
\begin{align*}
\frac{nq^2\delta E^2 \tau}{m} &= \frac{ B^2}{2\mu_0\tau} \\
\tau^2 &= \frac{ mB^2}{2\mu_0nq^2\delta E^2} \\
\tau &= \frac{B}{\delta E}\bigg(\frac{m}{2\mu_0nq^2}\bigg)^{1/2} \\ \\
c &= \frac{1}{\sqrt{\mu_0\epsilon_0}} \Rightarrow \frac{1}{\sqrt{\mu_0}} = \sqrt{\epsilon_0}c \\ \\
\Rightarrow \tau &= \frac{Bc}{\delta E}\bigg(\frac{\epsilon_0 m}{2nq^2}\bigg)^{1/2} \\ &= \frac{Bc}{\sqrt{2}\delta E \omega_p} \\ &= \frac{B \tau_{_E}}{\delta E} \equiv \tau_{_M}
\end{align*}

\subsection{Additional Onset Cases}
The same analysis as in section \ref{sec:reconnection} is applied here, only starting with $\ell_{_D}<\ell\sim\ell_{_{KS}}$.  In this case, $\ell>\ell_{_D}$ and therefore the density is depleting. Since $\ell\sim\ell_{_{KS}}$, the local electric field energy is not evolving.  
\begin{align}
\text{Onset Case V: } \ell_{_{D}}&<\ell \sim \ell_{_{KS}}:  \notag \\
|\nabla\cdot\vec{K}_{_D}|&<|\vec{J}\cdot\vec{E}|,|\nabla\cdot\vec{K}_{_E}|\sim|\nabla\cdot\vec{S}|  \\ \notag \\
\partial_t u_{_E} &=-\vec{J}\cdot\vec{E}- \nabla \cdot \vec{S} \sim 0 \\
\partial_t u_{_P} &\sim\vec{J}\cdot\vec{E}-\nabla \cdot \vec{K}_{_E} \sim 0 \\ 
\notag \\
\partial_t \ell_{_{KS}} &> 0 \Rightarrow \ell_{_{KS}} \rightarrow \infty  \\ 
\partial_t \ell_{_{D}} &\sim 0 \Rightarrow  \ell_{_{D}} \sim \text{const.} 
\end{align}
After some $\Delta t$, the new hierarchy of scales becomes $\ell_{_D}<\ell<\ell_{_{KS}}$.  Since $\ell<\ell_{_{KS}}$, the electric field energy is accumulating.  Since $\ell_{_D}<\ell_{_{KS}}$, the electric field accumulation dominates the evolution of $\ell_{_{KS}}$, causing it to decrease.     
\begin{align}
\text{Onset Case VI: } \ell_{_{D}}&<\ell < \ell_{_{KS}}:  \notag \\
|\nabla\cdot\vec{K}_{_D}|&<|\vec{J}\cdot\vec{E}|,|\nabla\cdot\vec{K}_{_E}|<|\nabla\cdot\vec{S}|  \\ \notag \\
\partial_t u_{_E} &\sim- \nabla \cdot \vec{S} > 0 \\
\partial_t u_{_P} &\sim\vec{J}\cdot\vec{E}-\nabla \cdot \vec{K}_{_E} \sim 0 \\ 
\notag \\
\partial_t \ell_{_{KS}} &< 0 \Rightarrow \ell \leftarrow \ell_{_{KS}}  \\ 
\partial_t \ell_{_{D}} &> 0 \Rightarrow  \ell_{_{D}}\rightarrow \ell 
\end{align}
If $\ell_{_{KS}}$ reaches $\ell$ first, then the system is back to case IV.  If $\ell_{_D}$ reaches $\ell$ first, then the critical scale order becomes $\ell_{_D}\sim\ell<\ell_{_{KS}}$.  The transport of plasma by $K_{_E}$ and $K_{_D}$ cancel out so the density doesn't evolve.  The electric field energy continues to accumulate since $\ell<\ell_{_{KS}}$, leading to $\ell_{_{KS}}$ decreasing and $\ell_{_D}$ increasing toward each other.
\begin{align}
\text{Onset Case VII: } \ell_{_{D}}&\sim\ell < \ell_{_{KS}}:  \notag \\
|\nabla\cdot\vec{K}_{_D}|&\sim|\vec{J}\cdot\vec{E}|,|\nabla\cdot\vec{K}_{_E}|<|\nabla\cdot\vec{S}|  \\ \notag \\
\partial_t u_{_E} &\sim- \nabla \cdot \vec{S} > 0 \\
\partial_t u_{_P} &\sim(\vec{J}\cdot\vec{E}-\nabla \cdot \vec{K}_{_E})- \nabla \cdot \vec{K}_{_D} \sim \vec{J}\cdot\vec{E} > 0 \\ 
\notag \\
\partial_t \ell_{_{KS}} &< 0 \Rightarrow \ell_{_D} \leftarrow \ell_{_{KS}}  \\ 
\partial_t \ell_{_{D}} &> 0 \Rightarrow  \ell_{_{D}}\rightarrow \ell_{_{KS}} 
\end{align}
The next immediate state is $\ell<\ell_{_D}<\ell_{_{KS}}$, in which case the density and electric field are increasing and therefore $\ell_{_{KS}}$ and $\ell_{_D}$ continue to approach each other, similar to the pre-onset case which didn't specify the order of $\ell_{_D}$ and $\ell_{_{KS}}$ with respect to each other.

\begin{figure}
    \centering
    \includegraphics[scale=0.45]{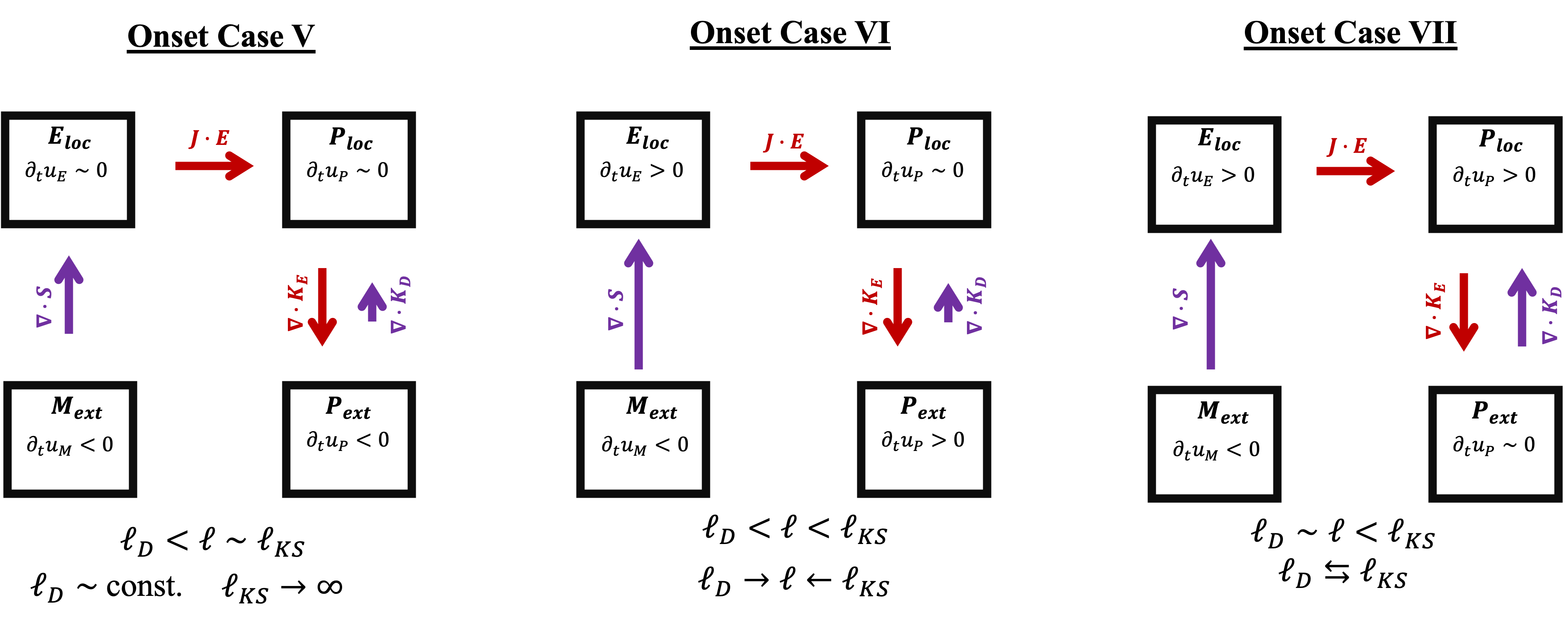}
    \caption{Additional onset cases and the resulting trends of the critical scales.}
    \label{fig:placeholder}
\end{figure}

\bibliographystyle{plain}
\bibliography{MTD}

\end{document}